\documentclass{agn_article}
\usepackage{agn_all}
\usepackage[hidelinks]{hyperref}
\usepackage{multirow}
\usepackage{booktabs}
\usepackage{float}
\usepackage{titlesec}
\usepackage{tocloft}

\titlespacing*{\paragraph}{0pt}{2ex plus 1ex minus .2ex}{1em}

\renewcommand{\stretch}{1.2}

\providecommand{\kappabar}{\overline{\kappa}}
\providecommand{\gammabar}{\overline{\gamma}}
\providecommand{\sigmahat}{\widehat{\sigma}}

\begin{document}
\title{A guide to enriched continuum modeling: advances on wave propagation in metamaterials through the relaxed micromorphic framework}

\defineauthor{hermann}{Svenja~Hermann}{bmsd}{svenja.hermann@tu-dortmund.de}
\defineauthor{sarhil}{Mohammad~Sarhil}{bmsd}{mohammad.sarhil@tu-dortmund.de}
\defineauthor{madeo}{Angela~Madeo}{bmsd}{angela.madeo@tu-dortmund.de}

\knownauthors[vossd]{vossd,hermann,sarhil,neff,madeo}
\maketitle

\vspace{-0.3cm}
\begin{abstract}
    \noindent
    This contribution provides a systematic framework for modeling the dispersive behavior of mechanical metamaterials by an enriched continua approach using the relaxed micromorphic model (RMM).
    We give the identification of its frequency-independent constitutive parameters from Bloch-Floquet dispersion data of the periodic metamaterial by combining analytical relations for characteristic features with a subsequent numerical optimization.
    Particular emphasis is placed on the underlying structure and limitations of the resulting dispersion relations, including their cut-offs, and asymptotic behavior.
    Finally, the transition from infinite periodic media to finite-size specimens is addressed, highlighting the role of the microscopic unit-cell cut and discussing null-Lagrangian contributions, effective interface forces, and surface Lagrangians as approaches to incorporate boundary effects.
\end{abstract}
{\textbf{Key words:} Metamaterials, metastructure,  inertia-augmented, relaxed micromorphic model, anisotropy, dispersion curves, band gap, parameters identification, generalized continua}
\\[.65em]
\noindent {\bf AMS 2010 subject classification: 74A30, 74A60, 74J10}

{\setlength{\cftbeforesecskip}{10pt}\parskip=-0.4mm\tableofcontents}

%
%
%
\section{Introduction}\label{voss.sec:1}

Metamaterials are architected materials whose effective mechanical properties are governed not only by the properties of their base material, but also, and often predominantly, by the geometry and characteristic length scales of their underlying microstructure. Through a suitable design of this microstructure, metamaterials can exhibit static and dynamic responses that are uncommon or even unattainable in conventional homogeneous materials. Examples include negative Poisson's ratios \cite{lakes1987foam}, unconventional coupling mechanisms between deformation modes \cite{frenzel2017three,rizzi2019identification}, frequency band-gaps \cite{liu2000locally,wang2014harnessing,bilal2018architected,celli2019bandgap}, wave focusing and cloaking \cite{guenneau2007acoustic,cummer2016controlling,buckmann2015mechanical,misseroni2016cymatics}, wave channeling \cite{kaina2017slow,tallarico2017edge}, and negative refraction \cite{zhu_negative_2014,kaina_negative_2015,willis_negative_2016}. These properties have stimulated considerable interest in the use of metamaterials for vibration mitigation, acoustic insulation, wave guiding, energy focusing and, more generally, for the control of elastic and acoustic waves.

Among these phenomena, the occurrence of frequency band-gaps is particularly relevant for engineering applications. Within a band-gap, propagating waves are strongly attenuated or completely inhibited, thus allowing specific frequency ranges to be filtered while others are transmitted. The position and width of such band-gaps are intrinsically linked to the microstructure. They depend on the mechanical properties of the constituent materials, on the characteristic size of the unit cell, and on its internal geometry. As a consequence, the design of a metamaterial for a prescribed operating frequency is inherently a multiscale problem: modifications performed at the scale of a single unit cell determine the wave-propagation properties observed at the scale of the complete structure.

The numerical description of this behavior is relatively straightforward when attention is restricted to a single periodic cell or to an infinite periodic medium. In this setting, Bloch-Floquet analysis combined with classical finite-element discretization provides an accurate description of the dispersion properties and allows the identification of propagating branches and frequency band-gaps. The situation becomes considerably more demanding when finite-size metamaterial structures are considered. A direct numerical model must explicitly resolve all geometrical details of every unit cell, including potentially slender ligaments, narrow gaps, or complex internal resonators. The resulting number of degrees of freedom rapidly becomes prohibitive as the number of unit cells increases, thereby limiting direct simulations of large-scale metastructures and making systematic design and optimization particularly expensive.

Classical homogenization provides an appealing way of overcoming this difficulty by replacing the heterogeneous material with an equivalent homogeneous continuum.
A broad mathematical and computational framework for such effective descriptions has been developed for static and quasi-static problems \cite{bensoussan2011asymptotic,sanchez1980non,allaire1992homogenization,milton2002theory,hill1963elastic,geers2010multi}.
Extensions of homogenization beyond a classical first-order Cauchy description can lead to higher-order or nonlocal effective continua \cite{pideri1997second,bouchitte2002homogenization,camar2003determination}, while dynamic homogenization approaches have been developed specifically to retain dispersive wave-propagation effects \cite{chen2001dispersive,craster2010high,bacigalupo2014second,willis2009exact,willis2011effective,srivastava2014limit,sridhar2018general}.
Nevertheless, when wavelengths become comparable to the characteristic size of the microstructure or when internal resonances are activated, a classical Cauchy continuum is generally unable to reproduce the strongly dispersive behavior of the underlying metamaterial.
In particular, the internal degrees of freedom responsible for optical branches and band-gap phenomena are lost in a purely classical effective description.

Enriched continuum theories offer a possible intermediate description between these two limiting approaches. Instead of explicitly resolving every geometrical detail of the microstructure, additional kinematic fields with micro-inertia terms are introduced at the continuum level. These additional degrees of freedom make it possible to retain information on the internal dynamics of the microstructure while describing the metamaterial as an effective homogeneous medium. In this way, enriched continua can reproduce dispersive phenomena that are inaccessible to a classical Cauchy description while, at the same time, considerably reducing the computational complexity associated with a direct microstructured model.

Different enriched continuum formulations vary in the choice of their additional kinematic fields, constitutive structure, and, most importantly, their kinetic and strain energy terms. These modeling choices directly determine the structure of the associated wave-propagation problem and therefore the number and qualitative features of the dispersion branches that the continuum can reproduce. The constitutive parameters subsequently determine the quantitative position and shape of these branches. This distinction between the \emph{structure of the model} and the \emph{identification of its parameters} will be an important aspect throughout this contribution.

Among the available enriched continuum theories, we focus on the relaxed micromorphic model as a representative framework for the effective description of mechanical metamaterials \cite{neff2014unifying,neff_relaxed_2015,madeo2015wave,d2020effective}. The purpose of the present contribution is not only to summarize its formulation, but to discuss how the different ingredients of the model are related to the wave-propagation characteristics of an underlying microstructured material and how they can be identified in practice.

We first consider the design of periodic metamaterials exhibiting band-gaps in the acoustic regime (our main application in practice) and discuss the influence of the constituent material, characteristic unit-cell size, and unit-cell geometry. We then introduce the enriched-continuum formulation used throughout and derive its governing equations and plane-wave dispersion problem. Particular attention is devoted to the mathematical structure of the resulting dispersion relations and to the role played by the different constitutive and inertial terms. Subsequently, a parameter-identification strategy based on the dispersion curves of the underlying microstructure is discussed. Finally, we consider the transition from infinite periodic media to finite-size metamaterial structures and outline both the capabilities and the remaining challenges of enriched-continuum descriptions for the design of large-scale metastructures.

%
%
%
\subsection{Metamaterial design}

The dynamic response of a mechanical metamaterial is primarily governed by three ingredients: the constitutive properties of the base material, the characteristic size of the unit cell, and the geometry of its internal architecture. All three influence the resulting dispersion properties and, in particular, the occurrence and position of acoustic band-gaps. In the present contribution, however, our main focus is on the role of the geometry. The base material is fixed and the influence of the unit-cell size can be treated separately through a scaling argument. This allows us to concentrate on the geometric mechanisms responsible for the emergence of band-gaps and, in particular, on the design choices required to obtain them in the acoustic regime.

\paragraph{Base material:}
Throughout, we assume the metamaterial to be made of polyethylene, as used for the additively manufactured specimens considered here. The corresponding material parameters are summarized in Table~\ref{voss.tab:baseMaterial}. Assuming isotropic linear elasticity, the bulk and shear moduli (in 3D) follow from
$K=\frac{E}{3\.(1-2\nu)}$ and $\mu=\frac{E}{2\.(1+\nu)}$, while the longitudinal and shear wave speeds are given by
$c_p=\sqrt{\frac{(3\.K+4\.\mu)}{3\.\rho}}$ and $c_s=\sqrt{\frac\mu\rho}$.

\begin{table}[h!]
    \centering
    \caption{Material parameters and corresponding elastic wave speeds of the polyethylene base material.}
    \label{voss.tab:baseMaterial}
    \renewcommand{\arraystretch}{\stretch}
    \begin{tabular}{llll}
    \toprule
    Quantity & Symbol & Value & Unit \\
    \midrule
    Mass density            & $\rho$   & $1220$           & $\mathrm{kg/m^3}$ \\
    Young's modulus         & $E$      & $1.45\times10^9$ & $\mathrm{Pa}$ \\
    Poisson's ratio         & $\nu$    & $0.40$            & -- \\
    Bulk modulus            & $K$ & $2.42\times10^9$ & $\mathrm{Pa}$ \\
    Shear modulus           & $\mu$    & $5.18\times10^8$ & $\mathrm{Pa}$ \\
    Longitudinal wave speed & $c_p$    & $1596$            & $\mathrm{m/s}$ \\
    Shear wave speed        & $c_s$    & $652$             & $\mathrm{m/s}$ \\
    \bottomrule
    \end{tabular}
\end{table}
The choice of the base material has a pronounced influence on the characteristic frequencies of the metamaterial, mainly through the associated stiffness, density, and associated wave speeds. Nevertheless, the purpose of the present contribution is not to optimize the base material itself. We therefore keep these properties fixed throughout and focus on the effects associated with the unit-cell size and, especially, its geometry.

\paragraph{Unit-cell size:}
The second key design variable is the characteristic size $a$ of the unit cell. For a fixed base material and fixed geometry, the dispersion response obeys a simple geometric scaling: characteristic wavelengths scale proportionally with $a$, whereas the corresponding frequencies scale inversely with $a$. Consequently, enlarging the unit cell shifts the relevant dispersion features towards lower frequencies, while reducing its size shifts them towards higher frequencies.

This scaling property allows the geometry and its effective continuum representation to be studied first for a reference unit-cell size. The target operating frequency can subsequently be shifted by uniformly rescaling the geometry without changing its topology. The absolute unit-cell size therefore does not have to be treated as an independent variable during the parameter-identification procedure considered later. In the relaxed micromorphic representation, the geometric scale is accounted for through the characteristic length $L_{\rm c}$, which we identify with the unit-cell size $a$ for simplicity.

\paragraph{Geometry:}
The geometry of the unit cell is the central design variable for the creation and tailoring of band-gaps.
Even comparatively simple microstructures can generate large frequency intervals in which wave propagation is prohibited.

\begin{figure}[h!]
    \centering
    \begin{minipage}[c]{0.63\textwidth}
        \centering
        \includegraphics[width=\linewidth,trim={2.6cm 2.4cm 2.7cm 2.7cm},clip]{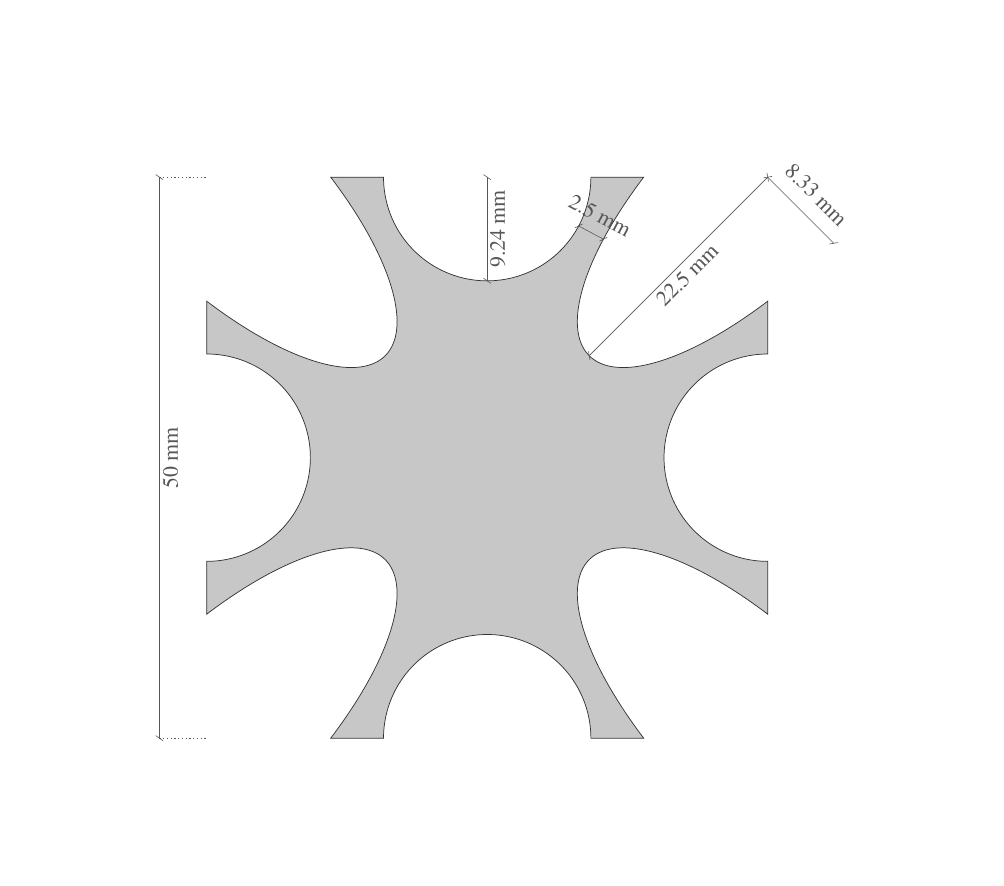}
        \smallskip
        \textbf{(a)} Unit-cell geometry
    \end{minipage}
    \hfill
    \begin{minipage}[c]{0.36\textwidth}
        \centering
        \includegraphics[width=\linewidth]{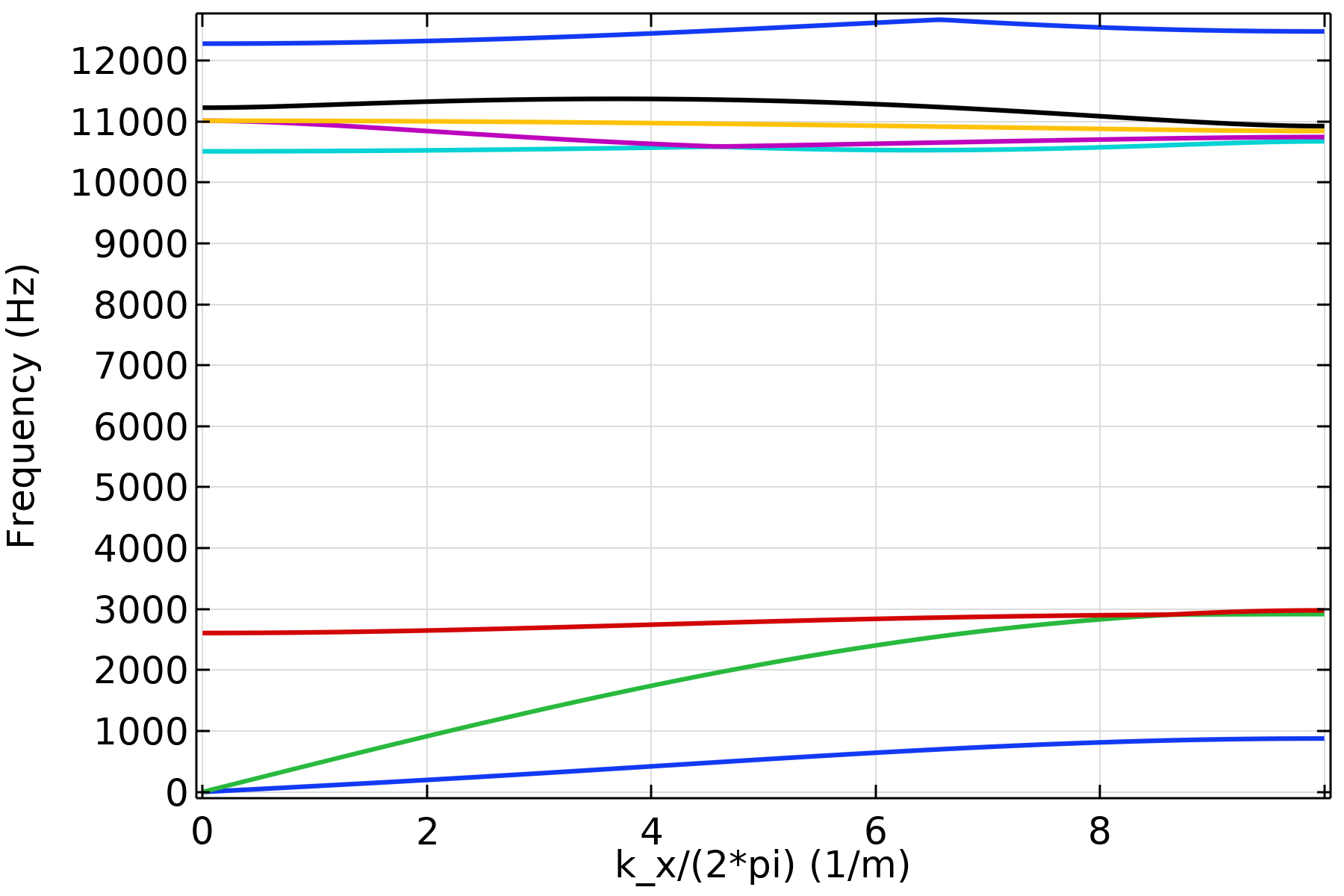}
        \smallskip
        \textbf{(b)} Dispersion diagram for $\phi=0^\circ$
        \vspace{1em}
        \includegraphics[width=\linewidth]{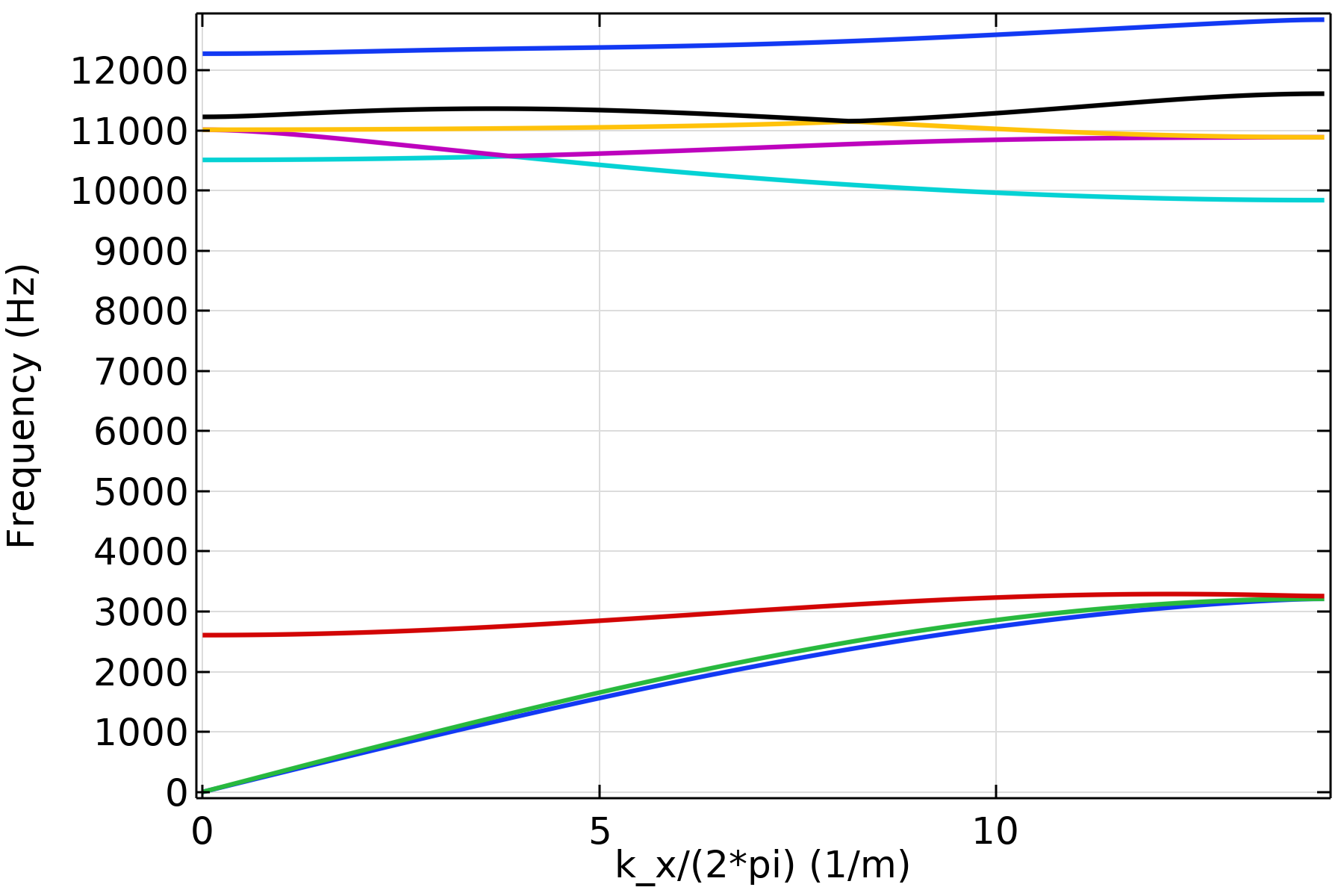}
        \smallskip
        \textbf{(c)} Dispersion diagram for $\phi=45^\circ$
    \end{minipage}
    \caption{Simple tetragonal metamaterial design and corresponding
    dispersion diagrams along the two principal symmetry directions.
    Even a comparatively simple unit-cell geometry can generate
    pronounced dispersive behavior and a wide frequency band-gap.}
    \label{voss.fig:voss_simple}
\end{figure}

Figure~\ref{voss.fig:voss_simple} shows a first example consisting of a tetragonal unit cell with circular and elliptical cuts together with its dispersion diagrams for propagation along the two principal symmetry directions $0^\circ$ and $45^\circ$.
The unit-cell size is $a=50\,\mathrm{mm}$, while the circular cut has radius $R_2\approx9.24\,\mathrm{mm}$ and the elliptical cuts are defined by the semi-axes $R_1=8\frac{1}{3}\,\mathrm{mm}$ and $R_4=22.5\,\mathrm{mm}$.
The geometry is chosen such that the minimum ligament between neighboring elliptical cuts is exactly $\delta=2.5\,\mathrm{mm}$.
Despite the geometric simplicity of the cell, its internal architecture is sufficient to produce pronounced dispersion and the opening of a remarkably wide band-gap.

Obtaining wide band-gaps at low frequencies (while keeping the same base material) is considerably more challenging. Low-frequency waves are associated with comparatively large wavelengths and, according to the scaling argument above, could in principle be targeted simply by increasing the unit-cell size. In practical applications, however, excessively large unit cells are usually undesirable or incompatible with the dimensions of the overall structure. The only alternative is therefore to modify the internal architecture for a constant unit-cell size such that additional compliant and inertial mechanisms are activated within a compact unit cell.

\begin{figure}[h!]
    \centering
    \begin{minipage}[c]{0.63\textwidth}
        \centering
        \includegraphics[width=\linewidth]{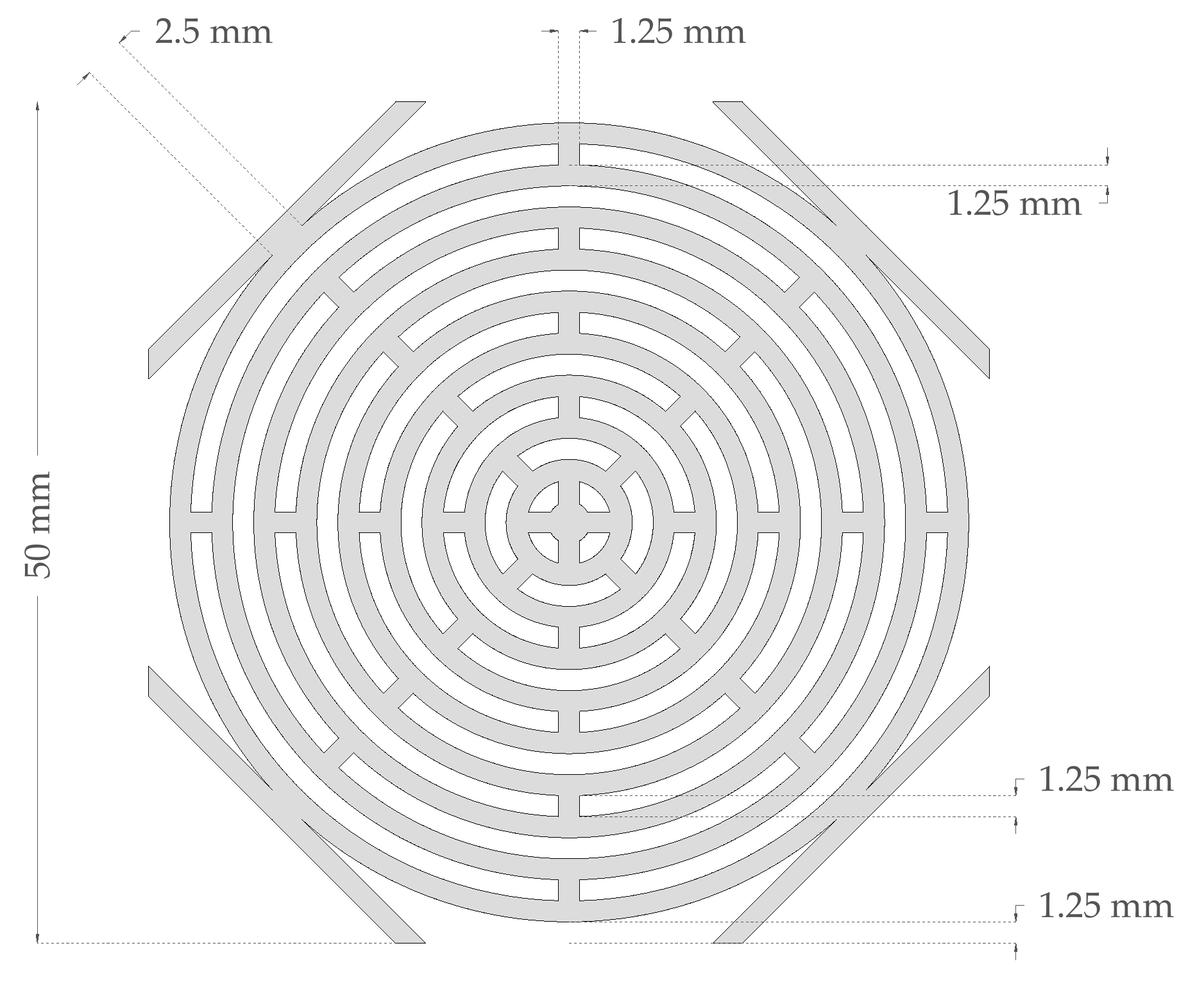}
        \smallskip
        \textbf{(a)} Labyrinthine unit-cell geometry
    \end{minipage}
    \hfill
    \begin{minipage}[c]{0.36\textwidth}
        \centering
        \includegraphics[width=\linewidth]{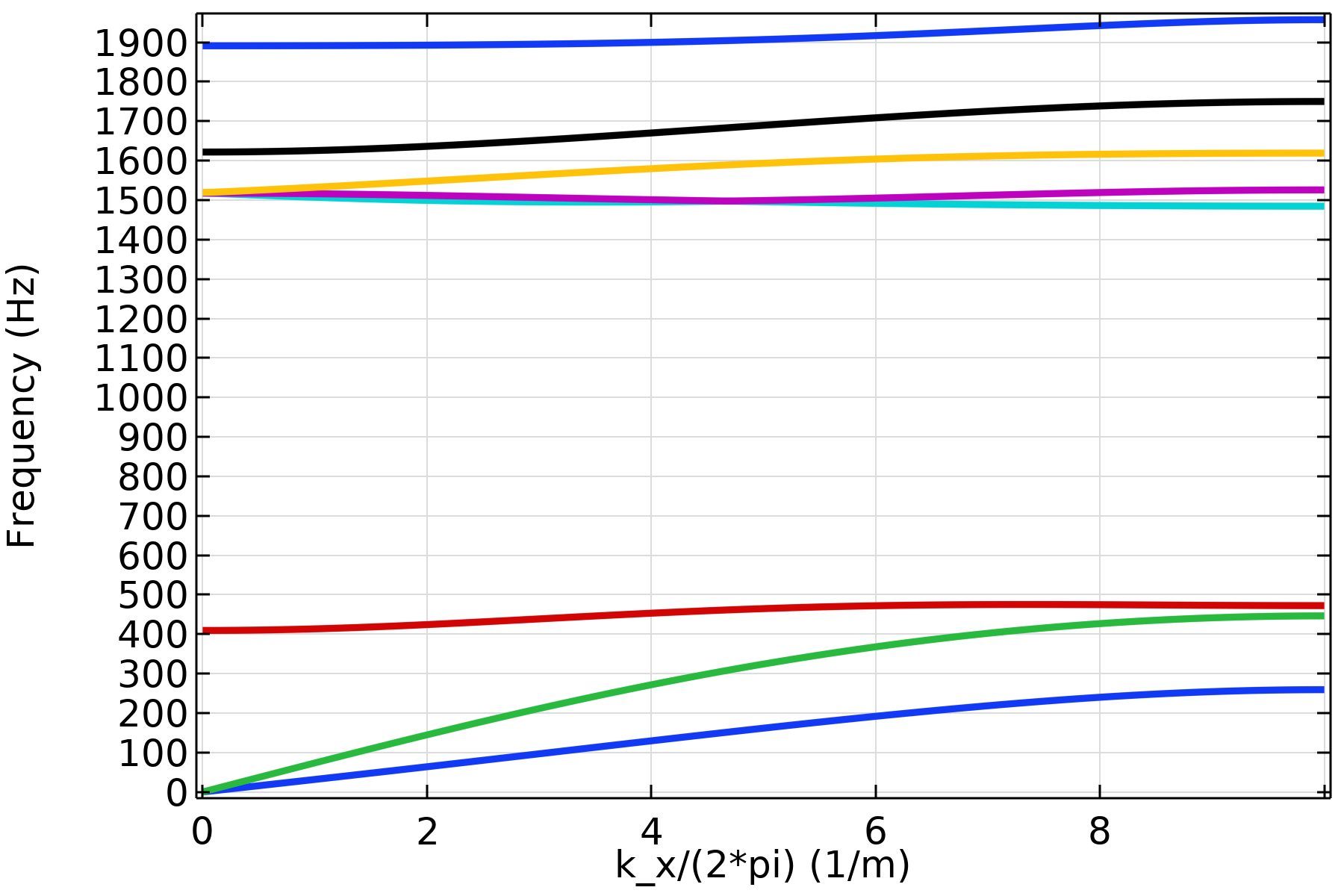}
        \smallskip
        \textbf{(b)} Dispersion diagram for $\phi=0^\circ$
        \vspace{1em}
        \includegraphics[width=\linewidth]{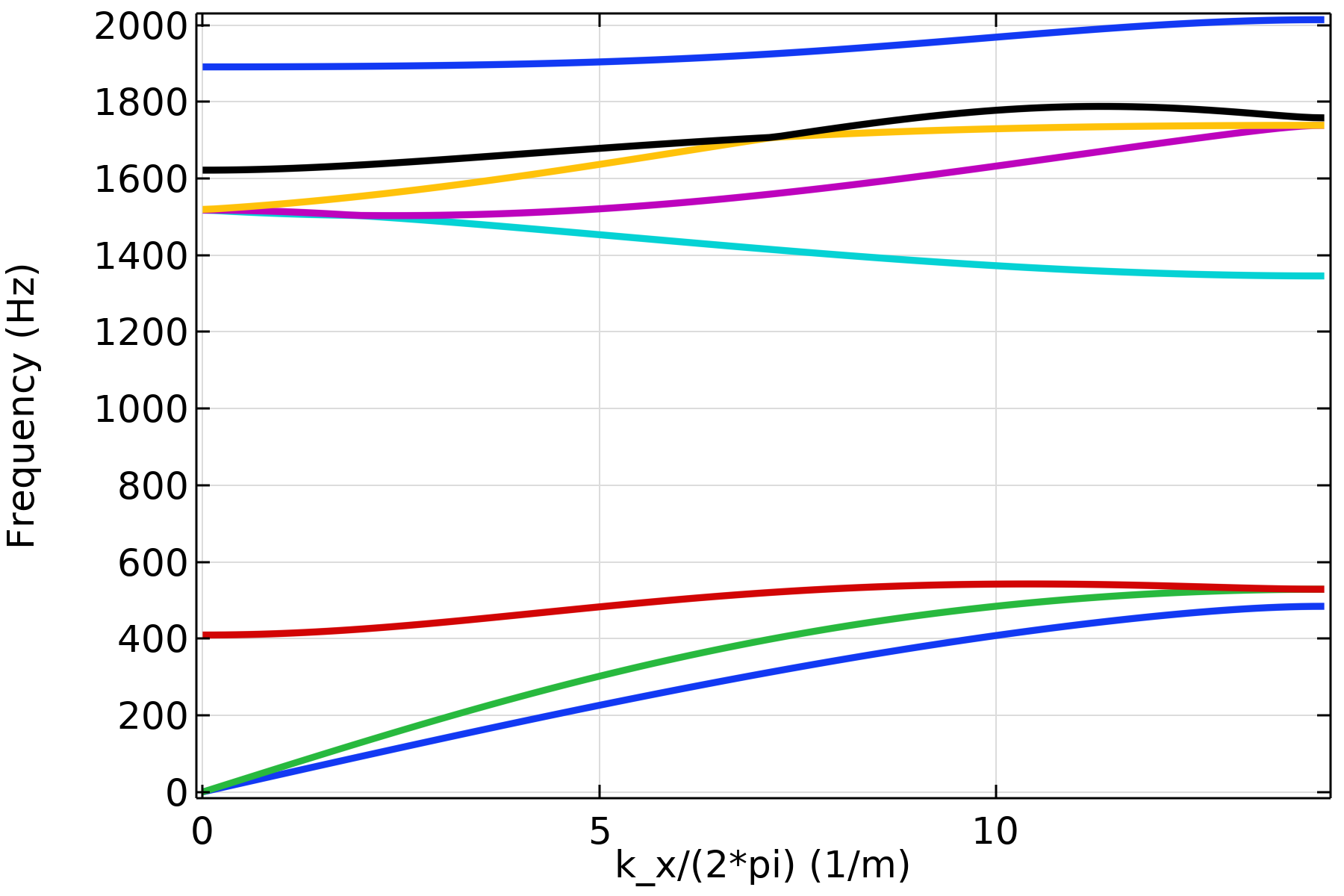}
        \smallskip
        \textbf{(c)} Dispersion diagram for $\phi=45^\circ$
    \end{minipage}
    \caption{Labyrinthine metamaterial introduced in
    \cite{voss2023modeling} for wave control in the low-frequency
    acoustic regime and corresponding dispersion diagrams along the
    two principal symmetry directions.}
    \label{voss.fig:voss_maze}
\end{figure}

One possible strategy is provided by labyrinthine geometries, in which long and slender material paths are folded into a comparatively small spatial domain, acting as a replacement for bigger unit cell size.
The labyrinthine unit cell shown in Figure~\ref{voss.fig:voss_maze} was initially designed in \cite{voss2023modeling,voss2023remarks} to generate a particularly wide band-gap at comparatively low frequencies (more than 6 times lower compared to the simpler geometry in Figure~\ref{voss.fig:voss_simple}) while maintaining a characteristic unit-cell size of only a few centimeters.
The geometry combines a strongly increased internal path length with a deliberate redistribution of mass and stiffness and therefore gives rise to a substantially richer dynamic response than the simpler geometry considered above.
Related labyrinthine and space-coiling concepts have been investigated for low-frequency sound control, impedance matching, wave manipulation, and broadband attenuation
\cite{xie2013tapered,frenzel2013three,krushynska2017spider,krushynska2018labyrinthine,liu2018fractal,kumar2020labyrinthine}.

The three design variables therefore play distinct but interconnected roles. The base material establishes the underlying elastic and inertial scales, the unit-cell size primarily determines the frequency scale, and the internal geometry controls the qualitative structure of the dispersion spectrum, including the occurrence, width, and position of band-gaps. In the following, the reference unit-cell is kept fixed (see Figure~\ref{voss.fig:voss_maze}), and we focus on how the dynamic properties generated by the microstructure can be represented by an effective enriched continuum.

%
%
%
\section{Enriched Continuum}\label{voss.sec:RMM}
Having characterized the dynamic response of the underlying microstructured material, the next step is to replace the explicitly re-solved unit-cell geometry by an effective continuum description. As discussed in the introduction, a classical Cauchy continuum is generally not sufficient for this purpose, since its kinematics do not retain the internal degrees of freedom responsible for the strongly dispersive behavior of metamaterials. Enriched continua overcome this limitation by introducing additional kinematic fields and associated elastic and inertial contributions at the macroscopic level.

In the following, we focus on one particular enriched continuum, namely the relaxed micromorphic model (RMM) \cite{neff2014unifying,madeo2015wave,d2020effective}. In addition to the macroscopic displacement field $u\in\R^3$, the RMM introduces an independent, non-symmetric micro-distortion tensor $P\in\R^{3\times3}$, which represents additional degrees of freedom associated with the motion and deformation of the underlying microstructure. The interaction between $u$ and $P$ with suitable elastic and micro-inertia contributions, enables the effective continuum to reproduce characteristic features of the dispersion spectrum of the corresponding microstructured material.

Although the subsequent derivations are presented for the RMM, the general procedure is not restricted to this particular model. Starting from a Lagrangian defined in terms of the chosen kinematic fields, the governing equations are always obtained through its first variation. A harmonic plane-wave ansatz then transforms the resulting system of partial differential equations into an algebraic system whose singularity condition defines the dispersion relation. The specific choice of kinematic variables and energy terms determines the mathematical structure of this dispersion problem, whereas the associated constitutive parameters determine its quantitative characteristics. Consequently, the same general workflow can be applied to other enriched continuum theories, although the resulting number and form of the dispersion branches may differ.

For the present analysis, we consider the inertia-augmented relaxed micromorphic formulation including the elastic and inertial terms required to describe the dispersion characteristics of the metamaterial over the frequency and wavenumber ranges of interest. 

\paragraph{Lagrangian:}
The constitutive structure of the relaxed micromorphic model is most conveniently introduced through the corresponding Lagrangian
\begin{equation}
    \mathcal{L}_{\rm m}\left(\nabla u,P,\Curl P,\dot{u},\nabla\dot{u},\dot{P},\Curl\dot{P}\right) 
	= \underbrace{J(\dot{u},\nabla\dot{u},\dot{P},\Curl\dot{P})}_{\text{kinetic energy}}
	- \underbrace{W(\nabla u,P,\Curl P)}_{\text{strain energy}}
\end{equation}
with the including terms \cite{voss2023modeling,madeo2015wave,d2020effective} defined as
\begin{align}
	W(\nabla u,P,\Curl P)&=\frac12\iprod{\mathbb{C}_{\rm e}\.\sym(\nabla u-P),\sym(\nabla u-P)}\\
	&\phantom{=}\ +\frac12\iprod{\mathbb{C}_{\rm c}\.\skew(\nabla u-P),\skew(\nabla u - P)} + \frac12\iprod{\mathbb{C}_{\rm micro}\.\sym P,\sym P}\notag\\
	&\phantom{=}\ +\frac12\iprod{\mathbb{L}_{\rm s}\.\sym\Curl P,\sym\Curl P} + \frac12\iprod{\mathbb{L}_{\rm a}\.\skew\Curl P,\skew\Curl P},\notag\\
	J(\dot{u},\nabla\dot{u},\dot{P},\Curl\dot{P})&=\frac12\.\rho\.\norm{\dot{u}}^2 + \frac12\iprod{\mathbb{J}_{\rm m}\.\sym\dot P,\sym\dot P}+\frac12\iprod{\mathbb{J}_{\rm c}\.\skew\dot P,\skew\dot P}\\
	&\phantom{=}\ +\frac12\iprod{\mathbb{T}_{\rm e}\.\sym\nabla \dot{u},\sym\nabla \dot{u}} + \frac12\iprod{\mathbb{T}_{\rm c}\.\skew\nabla \dot{u},\skew\nabla \dot{u}}\notag\\
	&\phantom{=}\ +\frac12\iprod{\mathbb{M}_{\rm s}\.\sym\Curl\dot P,\sym\Curl\dot P} + \frac12\iprod{\mathbb{M}_{\rm a}\.\skew\Curl\dot P,\skew\Curl\dot P},\notag
\end{align}
where $u\in\R^3$ is the macroscopic displacement field, $P\in\R^{3\times3}$ is the non-symmetric micro-distortion tensor, $\rho$ is the macroscopic apparent density, $\mathbb{J}_{\rm m}$, $\mathbb{J}_{\rm c}$, $\mathbb{T}_{\rm e}$, $\mathbb{T}_{\rm c}$, $\mathbb{M}_{\rm s}$, $\mathbb{M}_{\rm a}$, are 4th order micro-inertia tensors, and $\mathbb{C}_{\rm e}$, $\mathbb{C}_{\rm c}$, $\mathbb{C}_{\rm micro}$, $\mathbb{L}_{\rm s}$, and $\mathbb{L}_{\rm a}$ are 4th order elastic tensors.\footnote{
The tensors $\mathbb{J}_{\rm m}$, $\mathbb{T}_{\rm e}$, $\mathbb{C}_{\rm e}$, $\mathbb{C}_{\rm micro}$, $\mathbb{L}_{\rm s}$, and $\mathbb{M}_{\rm s}$ have a minor symmetry, i.e.\ $\mathbb{A}_{ijkl}=\mathbb{A}_{jikl}=\mathbb{A}_{ijlk}=\mathbb{A}_{jilk}$ while $\mathbb{J}_{\rm c}$, $\mathbb{T}_{\rm c}$, $\mathbb{C}_{\rm c}$, $\mathbb{L}_{\rm a}$, and $\mathbb{M}_{\rm a}$ has a minor anti-symmetry, i.e.\ $\mathbb{B}_{ijkl}=-\mathbb{B}_{jikl}$, $\mathbb{B}_{ijkl}=-\mathbb{B}_{ijlk}$, and $\mathbb{B}_{ijkl}=\mathbb{B}_{jilk}$. For instance, we can write $\iprod{\mathbb{J}_{\rm m}\.\sym\dot P,\sym\dot P}+\iprod{\mathbb{J}_{\rm c}\.\skew\dot P,\skew\dot P}=\iprod{(\mathbb{J}_{\rm m}+\mathbb{J}_{\rm c})\.\dot P,\dot P}$.}
We assume that the same class of symmetry applies both at the micro- and the macro-scale. Thus, all material tensors must match the symmetry of the associated unit cell which, in our case, is tetragonal symmetry, cf.\ Appendix \ref{voss.app:tensor}.
Overall, we end up with 19 frequency independent parameters:
\begin{itemize}
    \item the characteristic length $L_{\rm c}$ and the macroscopic apparent density $\rho$,
    \item the elastic parameters $\kappa_{\rm m},\mu_{\rm m},\mu_{\rm m}^*$ for the micro-scale, $\kappa_{\rm e},\mu_{\rm e},\mu_{\rm e}^*$ for meso-scale and $\mu_c$ for the skew-symmetric part,
    \item the inertia parameters $\kappa_\gamma,\gamma_1,\gamma_1^*,\gamma_2$ for $\dot P$ and $\overline{\kappa}_\gamma,\overline{\gamma}_1,\overline{\gamma}_1^*,\overline{\gamma_2}$ for $\nabla\dot u$, and
    \item the values $\alpha_1$ and $\beta_1$ corresponding to $\Curl P$ and $\Curl\dot P$, respectively.
\end{itemize}

\paragraph{Euler-Lagrange equation:}

We derive the Euler-Lagrange equations of the RMM from the stationarity of the action
\begin{equation}
    I(u,P)=\int_0^T\!\!\int_\Omega \mathcal{L}_{\rm m}\left(\nabla u,P,\Curl P,\dot{u},\nabla\dot{u},\dot{P},\Curl\dot{P}\right)\intd x\,\intd t\,.
\end{equation}
For arbitrary variations
\begin{equation}
    \delta u\colon\Omega\times[0,T]\to\R^3,
    \qquad\qquad
    \delta P\colon\Omega\times[0,T]\to\R^{3\times3},
\end{equation}
with $\delta u(\cdot,0)=\delta u(\cdot,T)=0$ and $\delta P(\cdot,0)=\delta P(\cdot,T)=0$,
we consider the perturbed fields
\begin{equation}
    u_\varepsilon=u+\varepsilon\.\delta u\,,
    \qquad\qquad
    P_\varepsilon=P+\varepsilon\.\delta P\,.
\end{equation}
The first variation of the action is defined by
\begin{equation}
    \delta I(u,P)[\delta u,\delta P]\coloneqq\left.\dd\varepsilon I(u_\varepsilon,P_\varepsilon)\right|_{\varepsilon=0}.
    \label{eq:stationarityCondition}
\end{equation}
For a fourth-order tensor $\mathbb A\in\R^{3\times3\times3\times3}$ with major symmetry and arbitrary second-order tensors $X,Y\in\R^{3\times3}$, we obtain
\begin{equation}
    \left.\dd\varepsilon\frac12\iprod{\mathbb A\.(X+\varepsilon Y),X+\varepsilon Y}\right|_{\varepsilon=0}
    =\frac12\left(\iprod{\mathbb A\.Y,X}+\iprod{\mathbb A\.X,Y}\right)
    =\iprod{\mathbb A\.X,Y}\,.
\end{equation}
Since all differential operators appearing in the Lagrangian are linear, we can apply the same argument to all quadratic terms in the relaxed micromorphic model
\begin{align}
    \delta I(u,P)[\delta u,\delta P]
    &=\int_0^T\!\!\int_\Omega
    \rho\iprod{\dot u,\delta\dot u}
    +\iprod{\mathbb J_{\rm m}\sym\dot P,\sym\delta\dot P}
    +\iprod{\mathbb J_{\rm c}\skew\dot P,\skew\delta\dot P
    }\notag\\
    &+\iprod{\mathbb T_{\rm e}\sym\nabla\dot u,\sym\nabla\delta\dot u}
    +\iprod{\mathbb T_{\rm c}\skew\nabla\dot u,\skew\nabla\delta\dot u
    }
    \label{voss.eq:first_variation_raw}\\
    &+\iprod{\mathbb M_{\rm s}\sym\Curl\dot P,\sym\Curl\delta\dot P}
    +\iprod{\mathbb M_{\rm a}\skew\Curl\dot P,\skew\Curl\delta\dot P}
    \notag\\
    &-\iprod{\mathbb C_{\rm e}\sym(\nabla u-P),\sym(\nabla\delta u-\delta P)}
    -\iprod{\mathbb C_{\rm c}\skew(\nabla u-P),\skew(\nabla\delta u-\delta P)}
    \notag\\
    &-\iprod{\mathbb C_{\rm micro}\sym P,\sym\delta P}\notag\\
    &-\iprod{\mathbb L_{\rm s}\sym\Curl P,\sym\Curl\delta P}
    -\iprod{\mathbb L_{\rm a}\skew\Curl P,\skew\Curl\delta P}
    \;\intd x\,\intd t\,.\notag
\end{align}
To simplify the notation, we introduce
\begin{align}
	\widetilde{\sigma}&\coloneqq
	\mathbb{C}_{\rm e}\.\sym(\nabla u-P) + \mathbb{C}_{\rm c}\.\skew(\nabla u-P)\,,
	&\sigmahat&\coloneqq
	\mathbb{T}_{\rm e}\.\sym\nabla\ddot{u} + \mathbb{T}_{\rm c}\.\skew\nabla\ddot{u}\,,\notag\\
	\overline\sigma&\coloneqq
	\mathbb{J}_{\rm m}\.\sym\ddot{P} + \mathbb{J}_{\rm c}\.\skew\ddot{P}\,,
	&s&\coloneqq
	\mathbb{C}_{\rm micro}\. \sym P \,,\\
	m&\coloneqq
	\mathbb{L}_{\rm s}\.\sym\Curl P + \mathbb{L}_{\rm a}\.\skew\Curl P\,,
	&\widehat m&\coloneqq
	\mathbb{M}_{\rm s}\.\sym\Curl\ddot{P} + \mathbb{M}_{\rm a}\.\skew\Curl\ddot{P} \,.\notag
\end{align}
For $S=S^T\in\R^{3\times3}$ (symmetric) and $A=-A^T\in\R^{3\times3}$ (skew-symmetric), it holds
\begin{equation}
    \iprod{S,\sym X}=\iprod{S,X}
    \qquad\text{and}\qquad
    \iprod{A,\skew X}=\iprod{A,X}\,.
\end{equation}
Now, the first variation \eqref{voss.eq:first_variation_raw} can be written in the compact form
\begin{align}
    \delta I(u,P)[\delta u,\delta P]
    =\int_0^T\!\!\int_\Omega
    &\rho\iprod{\dot u,\delta\dot u}
    +\iprod{\mathbb{J}_{\rm m}\.\sym\dot P+\mathbb{J}_{\rm c}\.\skew\dot P,\delta\dot P}\notag\\
    &+\iprod{\mathbb{T}_{\rm e}\.\sym\nabla\dot u+\mathbb{T}_{\rm c}\.\skew\nabla\dot u,\nabla\delta\dot u}
    \label{voss.eq:first_variation_compact}\\
    &+\iprod{\mathbb{M}_{\rm s}\.\sym\Curl\dot P+\mathbb{M}_{\rm a}\.\skew\Curl\dot P,\Curl\delta\dot P}\notag\\
    &-\iprod{\widetilde{\sigma},\nabla\delta u-\delta P}
    -\iprod{s,\delta P}
    -\iprod{m,\Curl\delta P}
    \;\dx\,\dt\,.\notag
\end{align}
We first integrate the kinetic contributions by parts with respect
to time. In general,
\begin{equation}
    \int_0^T \iprod{A,\dot B}\;\dt
    =\left.\iprod{A,B}\right|_0^T
    -\int_0^T \iprod{\dot A,B}\;\dt\,.
    \label{voss.eq:partial_integration_time}
\end{equation}
The first part is always zero because the variations $\delta u$ and $\delta P$ are assumed to vanish at the initial and final time.
Consequently,
\begin{align}
    \int_0^T
    \rho\iprod{\dot u,\delta\dot u}\;\dt
    &=-\int_0^T\rho\iprod{\ddot u,\delta u}\;\dt\,,\notag\\
    \int_0^T\iprod{\mathbb{J}_{\rm m}\.\sym\dot P+\mathbb{J}_{\rm c}\.\skew\dot P,\delta\dot P}\;\dt
    &=-\int_0^T\iprod{\overline{\sigma},\delta P}\;\dt\,,\\
    \int_0^T\iprod{\mathbb{T}_{\rm e}\.\sym\nabla\dot u+\mathbb{T}_{\rm c}\.\skew\nabla\dot u,\nabla\delta\dot u}\;\dt
    &=-\int_0^T\iprod{\widehat{\sigma},\nabla\delta u}\;\dt\,,\notag\\
    \int_0^T\iprod{\mathbb{M}_{\rm s}\.\sym\Curl\dot P+\mathbb{M}_{\rm a}\.\skew\Curl\dot P,\Curl\delta\dot P}\;\dt
    &=-\int_0^T\iprod{\widehat m,\Curl\delta P}\;\dt\,.\notag
\end{align}
Substitution into the first variation \eqref{voss.eq:first_variation_compact} gives
\begin{align}
    \delta I(u,P)[\delta u,\delta P]
    =\int_0^T\!\!\int_\Omega
    &-\rho\iprod{\ddot u,\delta u}
    -\iprod{\widetilde{\sigma}+\widehat{\sigma},\nabla\delta u}\notag\\
    &+\iprod{\widetilde{\sigma}-s-\overline{\sigma},\delta P}
    -\iprod{m+\widehat m,\Curl\delta P}\;\dx\,\dt\,.
    \label{voss.eq:first_variation_time}
\end{align}
It remains to remove the spatial derivatives of the variations $\nabla\delta u$ and $\Curl\delta P$.
For a sufficiently regular tensor field $A$, integration by parts
and Gauss' theorem yield
\begin{equation}
    -\int_\Omega\iprod{A,\nabla\delta u}\;\dx
    =\int_\Omega\iprod{\Div A,\delta u}\;\dx
    -\int_{\partial\Omega}\iprod{A\.n,\delta u}\;\intd{S}\,,
    \label{voss.eq:partial_integration_div}
\end{equation}
where $n$ denotes the outward unit normal to $\partial\Omega$.
For the row-wise Curl operator, we use
\begin{equation}
    \int_\Omega\iprod{A,\Curl\delta P}\;\dx
    =-\int_{\partial\Omega}\iprod{A,\delta P\times n}\;\intd{S}
    +\int_\Omega\iprod{\Curl A,\delta P}\;\dx\,.
    \label{voss.eq:partial_integration_curl}
\end{equation}
Using $\iprod{A,\delta P\times n}=-\iprod{A\times n,\delta P}$, the boundary contribution can equivalently be written directly in
terms of $\delta P$.
Applying the last two terms \eqref{voss.eq:partial_integration_div} and \eqref{voss.eq:partial_integration_curl} to the first variation \eqref{voss.eq:first_variation_time}, we finally obtain
\begin{align}
    \delta I(u,P)[\delta u,\delta P]
    =\int_0^T\!\!\int_\Omega
    &\iprod{\Div(\widetilde{\sigma}+\widehat{\sigma})-\rho\ddot u,\delta u}\notag\\
    +&\iprod{\widetilde{\sigma}-s-\overline{\sigma}-\Curl(m+\widehat m),\delta P}\;\dx\,\dt
    \label{voss.eq:first_variation_final}\\
    -\int_0^T\!\!\int_{\partial\Omega}&\iprod{(\widetilde{\sigma}+\widehat{\sigma})\.n,\delta u}
    +\iprod{(m+\widehat m)\times n,\delta P}\;\intd{S}\,\dt\,.\notag
\end{align}
For variations with compact support in $\Omega$, the two boundary terms
vanish. With the fundamental lemma of the calculus of variations, the stationarity condition \eqref{eq:stationarityCondition} for all independent variations $\delta u$ and $\delta P$, gives the Euler-Lagrange equations
\begin{equation}
    \Div(\widetilde{\sigma}+\widehat{\sigma})-\rho\ddot u=0\,,
    \qquad\qquad
    \widetilde{\sigma}-s-\overline{\sigma}-\Curl(m+\widehat m)=0\,.
\end{equation}
or, equivalently
\begin{equation}
    \rho\.\ddot{u}-\Div\widehat{\sigma}
    =\Div\widetilde{\sigma}\,,
    \qquad\qquad
    \overline{\sigma}
    =\widetilde{\sigma}-s-\Curl m-\Curl\widehat m\,.
    \label{voss.eq:EL}
\end{equation}
If the variations are not prescribed on the boundary, the boundary
terms in \eqref{voss.eq:first_variation_final} additionally provide
the natural Neumann-type boundary conditions
\begin{equation}
    (\widetilde{\sigma}+\widehat{\sigma})\.n=0\,,
    \qquad(m+\widehat m)\times n=0
    \qquad\text{on}\quad\partial\Omega\,.
\end{equation}
The well posedness of this dynamical problem under natural assumptions has been proven in \cite{owczarek2019nonstandard}.

%
%
%
\subsection{Plane waves}
As it is known in the context of dynamical analysis, a particular class of solutions of the system of partial differential equations \eqref{voss.eq:EL} can be found considering the monochromatic plane wave form for the involved kinematic fields, i.e.
\begin{equation}
	u(x,t)=\uhat \. e^{i\left( \iprod{\boldsymbol k,x} - \omega \. t \right)},\qquad P(x,t)=\Phat \. e^{i\left( \iprod{\boldsymbol k,x} - \omega \. t \right)},\qquad \boldsymbol k = k\.\khat\,,
	\label{voss.eq:time_harm_ansatz}
\end{equation}
where $\uhat\in\C^3$ and $\Phat\in\C^{3\times3}$ are the so-called polarization vectors, $\khat\in\R^3$ is the direction of wave propagation with $\norm{\khat}=1$ and wavenumber $k\in\R$.

The plane-wave ansatz transforms the PDE system \eqref{voss.eq:EL} into an algebraic system because spatial and temporal differentiation acts only on the common exponential factor. Indeed,
\begin{equation}
    \nabla u
    =i\.\uhat\otimes\boldsymbol{k}\.
    e^{i(\iprod{\boldsymbol{k},x}-\omega t)},
    \qquad\qquad
    \ddot u
    =-\omega^2\uhat\.
    e^{i(\iprod{\boldsymbol{k},x}-\omega t)},
\end{equation}
and analogously
\begin{equation}
    \nabla P
    =i\.\Phat\otimes\boldsymbol{k}\.e^{i(\iprod{\boldsymbol{k},x}-\omega t)},
    \qquad\qquad
    \ddot P
    =-\omega^2\Phat\.e^{i(\iprod{\boldsymbol{k},x}-\omega t)}.
\end{equation}
Hence, every spatial derivative introduces a factor proportional to $i\.\boldsymbol{k}$ and every second temporal derivative a factor $-\omega^2$, while all terms in the Euler-Lagrange equations retain the same factor $e^{i(\iprod{\boldsymbol{k},x}-\omega t)}$.
Since this exponential can never be zero, it can be divided out from the complete system.
The PDE-system therefore reduces to the homogeneous algebraic problem
\begin{equation}
    A(\omega,\boldsymbol{k})\.v=0\,,
    \qquad\qquad
    v=(\uhat,\Phat)\,,
    \label{voss.eq:eigenProb}
\end{equation}
where $A(\omega,\boldsymbol{k})\in\C^{12\times12}$ is the \emph{acoustic tensor}.
We will be interested only in the study of the wave propagation with $\khat$ in the plane $(x_1, x_2, 0)$ which allows us to write $\khat=(\cos\phi,\sin\phi, 0)$, where $\phi$ is the angle of propagation.
In this way, setting the amplitudes out of the plane equal to zero, the vector of the unknown amplitudes $v$ reduces to $v=(\uhat_1,\Phat_{11},\Phat_{22},\uhat_2,\Phat_{12},\Phat_{21})$ and the acoustic tensor $A(\omega,k,\phi)\in\C^{6\times6}$.

The non-trivial solutions of the system \eqref{voss.eq:eigenProb} are obtained when $A$ is singular, i.e.\ when $\det A=0$, which provides relations between $\omega=\omega(k,\phi)$ (or $k=k(\omega,\phi)$), the so-called \emph{dispersion curves}. We summarize:
\begin{itemize}
    \item The harmonic plane-wave ansatz reduces the governing equations into a homogeneous algebraic linear system $A\.v=0$.
    \item The dimension of unknown amplitudes $v$ is given by the number of kinematic fields, here $u$ and $P$ under the plane-strain assumption.
    \item The dispersion curves are \textbf{polynomial roots} $\det A=0$ with respect to the frequency $\omega$ or wavenumber $k$.
    \item The material parameters determine the final shape of these dispersion curves.
\end{itemize}
We want to emphasize that the maximal number of dispersion branches and of admissible wavenumber solutions is determined by the polynomial degree of $\det A(\omega,k,\phi)$. Thereby, the main mechanism in the transformation from \eqref{voss.eq:EL} to \eqref{voss.eq:eigenProb} are the number of derivatives occurring in our formulation, obtaining $k$ for each spacial derivative and $\omega$ for each temporal derivative of \eqref{voss.eq:time_harm_ansatz}.
For the RMM, we can write
\begin{equation}
    A=A_0+A_1\.k+A_2\.k^2+A_3\.\omega^2+A_4\.\omega^2\.k^2
\end{equation}
with the material parameter dependencies
\begin{align}
    A_0&=A_0(\C_{\rm micro},\C_{\rm e},\C_{\rm c})\,,&
    A_1&=A_1(\C_{\rm e},\C_{\rm c})\,,&
    A_2&=A_2(\C_{\rm e},\C_{\rm c},\mathbb{L}_{\rm s},\mathbb{L}_{\rm a})\,,\notag\\
    A_3&=A_3(\rho,\mathbb{J}_{\rm m},\mathbb{J}_{\rm c})\,,&
    A_4&=A_4(\mathbb{T}_{\rm e},\mathbb{T}_{\rm c},\mathbb{M}_{\rm s},\mathbb{M}_{\rm a})\,.
\end{align}
For a general propagation direction $\phi$, the six in-plane amplitudes remain coupled and the dispersion relation follows from the determinant of the full matrix $A(\omega,k,\phi)\in\C^{6\times6}$.
In the following, we focus on propagation along the two symmetry directions $\phi=0^\circ$ and $\phi=45^\circ$ of the tetragonal material.
Along these directions, the in-plane system decouples into two independent
$3\times3$ blocks,
\begin{equation}
    \det A(\omega,k,\phi)
    =\det B_{\rm p}(\omega,k,\phi)\.\det B_{\rm s}(\omega,k,\phi)\,,
    \qquad
    \phi\in\{0^\circ,45^\circ\}\,,
\end{equation}
where $B_{\rm p}$ acts on the pressure-wave amplitudes and $B_{\rm s}$ on the shear-wave amplitudes.
Accordingly, the pressure and shear dispersion branches can be analyzed independently.
When also rotating the propagation basis the two blocks corresponds to the pressure wave components $(\uhat_1,\Phat_{11},\Phat_{22})$ and shear waves components $(\uhat_2,\Phat_{12},\Phat_{21})$, respectively.
For each of these decoupled problems \cite[Appendix D]{voss2023modeling}, the dispersion relation takes the form
\begin{equation}
    c_0\.k^2+c_1\.k^4-(c_2+c_3\.k^2+c_4\.k^4)\,\omega^2+(c_5+c_6\.k^2+c_7\.k^4)\,\omega^4-(c_8+c_9\,k^2+c_{10}\.k^4)\,\omega^6=0\,.
    \label{voss.eq:dispRelationAll}
\end{equation}
For every prescribed wavenumber $k$, \eqref{voss.eq:dispRelationAll} is a cubic polynomial in $\omega^2$.
Consequently, each $3\times3$ block gives rise to three dispersion branches, corresponding in the present setting to one acoustic and two optical modes.
The decoupled pressure and shear problems therefore provide three pressure and three shear branches, respectively, i.e.\ six in-plane dispersion curves in total.
This number is fixed by the structure of the model and cannot be changed by the subsequent parameter identification.
We return to this structural restriction in Section~\ref{voss.sec:conceptualLimitations}, where the dispersion relation is analyzed in more detail.

Again, we want to emphasize, that here and in the following, the coefficients $c_i$ are understood to depend on the considered wave type and direction of propagation for each of the four decoupled cases $(\mathrm{p},0^\circ)$, $(\mathrm{s},0^\circ)$, $(\mathrm{p},45^\circ)$, and $(\mathrm{s},45^\circ)$.

%
%
%
\subsection{Null-Lagrangian of the gradient micro-inertia}\label{voss.sec:kernel}
There is one additional observation in the identification of the gradient micro-inertia parameters $\mathbb{T}_{\rm e}$ and $\mathbb{T}_{\rm c}$ of $\nabla\dot u$.
Following \cite{erel2025null}, we introduce the linear combinations
\begin{equation}
    \begin{array}{l}
         \overline{\kappa}_1\coloneqq\overline{\kappa}_\gamma+\overline{\gamma}_1\,,\\
         \overline{\kappa}_2\coloneqq\overline{\gamma}_1^*+\overline{\gamma}_2\,,\\
         \overline{\kappa}_3\coloneqq\overline{\kappa}_\gamma-\overline{\gamma}_1+\overline{\gamma}_1^*-\overline{\gamma}_2\,,\\
         \overline{\kappa}_4\coloneqq\overline{\kappa}_\gamma-\overline{\gamma}_1-\overline{\gamma}_1^*+\overline{\gamma}_2\,.
    \end{array}
    \qquad\implies\qquad
    \matr{\overline{\kappa}_1\\ \overline{\kappa}_2\\ \overline{\kappa}_3\\ \overline{\kappa}_4}
    =\matr{1&1&0&0\\ 0&0&1&1\\ 1&-1&1&-1\\ 1&-1&-1&1}
    \matr{\overline{\kappa}_\gamma\\ \overline{\gamma}_1\\ \overline{\gamma}_1^*\\ \overline{\gamma}_2}.
    \label{voss.eq:gradient_inertia_combinations}
\end{equation}
Under plane strain, the gradient-inertia stress reads
\begin{equation}
    \widehat{\sigma}
    =\rho L_{\rm c}^2
    \matr{
        (\overline{\kappa}_\gamma+\overline{\gamma}_1)\ddot u_{1,1}+(\overline{\kappa}_\gamma-\overline{\gamma}_1)\ddot u_{2,2}
        &
        (\overline{\gamma}_1^*+\overline{\gamma}_2)\ddot u_{1,2}+(\overline{\gamma}_1^*-\overline{\gamma}_2)\ddot u_{2,1}
        \\
        (\overline{\gamma}_1^*-\overline{\gamma}_2)\ddot u_{1,2}+(\overline{\gamma}_1^*+\overline{\gamma}_2)\ddot u_{2,1}
        &
        (\overline{\kappa}_\gamma-\overline{\gamma}_1)\ddot u_{1,1}+(\overline{\kappa}_\gamma+\overline{\gamma}_1)\ddot u_{2,2}
    }.
\end{equation}
Using the parameter transformation \eqref{voss.eq:gradient_inertia_combinations}, this can equivalently be written as
\begin{equation}
    \widehat{\sigma}
    =\rho L_{\rm c}^2
    \matr{
        \overline{\kappa}_1\ddot u_{1,1}+\frac12\overline{\kappa}_3\ddot u_{2,2}
        &
        \overline{\kappa}_2\ddot u_{1,2}+\frac12\overline{\kappa}_3\ddot u_{2,1}
        \\
        \overline{\kappa}_2\ddot u_{2,1}+\frac12\overline{\kappa}_3\ddot u_{1,2}
        &
        \overline{\kappa}_1\ddot u_{2,2}+\frac12\overline{\kappa}_3\ddot u_{1,1}
    }
    +\frac{\rho L_{\rm c}^2}{2}\overline{\kappa}_4\dot{\sigma}^{\,0}\,,
\end{equation}
where
\begin{equation}
    \sigma^{0}
    \coloneqq\matr{\dot u_{2,2} & -\dot u_{2,1}\\ -\dot u_{1,2} & \dot u_{1,1}}
    =\Cof\nabla\dot u
    \qquad\implies\qquad
    \Div\dot{\sigma}^{\,0}
    =\matr{\ddot u_{2,21}-\ddot u_{2,12}\\ -\ddot u_{1,21}+\ddot u_{1,12}}
    =0\,.
\end{equation}
Therefore, the contribution of the gradient micro-inertia to the bulk equation is
\begin{equation}
    \Div\widehat{\sigma}
    = \rho L_{\rm c}^2\matr{\overline{\kappa}_1\ddot u_{1,11}+\overline{\kappa}_2\ddot u_{1,22}+\overline{\kappa}_3\ddot u_{2,12}\\ \overline{\kappa}_2\ddot u_{2,11}+\overline{\kappa}_1\ddot u_{2,22}+\overline{\kappa}_3\ddot u_{1,12}}.
    \label{voss.eq:gradient_inertia_bulk}
\end{equation}
Hence, the bulk operator, and therefore also the dispersion relations, depend only on $\overline{\kappa}_1$, $\overline{\kappa}_2$, and $\overline{\kappa}_3$, but not on $\overline{\kappa}_4$.

The parameter $\overline{\kappa}_4$ thus parametrizes a null-Lagrangian contribution: it is invisible to the bulk equations and consequently cannot be identified from Bloch-Floquet dispersion data.
Nevertheless, $\widehat{\sigma}^{\,0}$ contributes to the natural boundary traction,
\begin{equation}
    t_{\rm ker}
    =\widehat{\sigma}^{\,0}\.n
    =\frac{\rho L_{\rm c}^2}{2}
    \overline{\kappa}_4\.\dot{\sigma}^{\,0}\.n\,.
    \label{voss.eq:kernel_traction}
\end{equation}
Thus, different gradient micro-inertia tensors can describe exactly the same infinite bulk material while giving rise to different responses for finite-size specimens.
%
%
%
\subsection{Analytical parameter identification}
Overall, the relaxed micromorphic model has 19 parameters in the plane strain tetragonal symmetry case
\begin{align}
	&\rho\,,\;L_{\rm c}\,,\quad
	\mathbb{C}_{\rm micro}\sim\{\kappa_{\rm m},\mu_{\rm m},\mu_{\rm m}^*\}\,,\quad
	\mathbb{C}_{\rm c}\sim\{\mu_{\rm c}\}\,,\quad
	\mathbb{C}_{\rm e}\sim\{\kappa_{\rm e},\mu_{\rm e},\mu_{\rm e}^*\}\,,\quad
	\mathbb{L}_{\rm s},\mathbb{L}_{\rm a}\sim\{\alpha_1\}\,,\notag\\
	&\mathbb{J}_{\rm m}\sim\{\kappa_\gamma,\gamma_1,\gamma_1^*\}\,,\quad
	\mathbb{J}_{\rm c}\sim\{\gamma_2\}\,,\quad
	\mathbb{T}_{\rm e}\sim\{\overline{\kappa}_\gamma,\overline{\gamma}_1,\overline{\gamma}_1^*\}\,,\quad
	\mathbb{T}_{\rm c}\sim\{\overline{\gamma}_2\}\,,\quad
	\mathbb{M}_{\rm s},\mathbb{M}_{\rm a}\sim\{\beta_1\}\,.
\end{align}
Before fitting the complete dispersion curves numerically, a considerable number of parameters can be determined, or eliminated, analytically from geometric information and characteristic features of the Bloch-Floquet dispersion curves.
We proceed in three steps.

\subparagraph{Characteristic length and apparent density:}
We identify the characteristic length of the relaxed micromorphic model with the characteristic size of the periodic unit cell, i.e.\ $L_{\rm c}=a$.
This choice ensures consistency with a uniform rescaling of the microstructure: if the unit-cell geometry is scaled by a factor $\lambda>0$, both the characteristic frequencies and wavenumbers scale with $1/\lambda$, while the dimensionless constitutive parameters remain unchanged \cite{voss2023modeling}.

The macroscopic apparent density follows directly from the mass contained in one unit cell.
For the present plane-strain setting with a homogeneous base material of density $\rho_{\rm B}$, it is given by
\begin{equation}
    \rho=\rho_{\rm B}\.\frac{A_{\rm solid}}{A_{\rm cell}}
    =\rho_{\rm B}\left(1-\frac{A_{\rm void}}{A_{\rm cell}}\right),
    \label{voss.eq:apparent_density}
\end{equation}
where $A_{\rm cell}$, $A_{\rm solid}$, and $A_{\rm void}$ are the total, solid, and void area of the unit cell, respectively.

\subparagraph{Macroscopic elastic limit:}
The effective elastic response in the long-wavelength limit is described by the macroscopic elasticity tensor $\mathbb{C}_{\rm macro}$ which is also known as the effective stiffness tensor uniquely determined from classical periodic homogenization \cite{neff2020identification}.
For the tetragonal plane-strain setting considered here, it is parametrized by
\begin{equation}
    \mathbb{C}_{\rm macro}\sim\left\{\kappa_{\rm M},\mu_{\rm M},\mu_{\rm M}^{*}\right\}.
\end{equation}
Alternatively, these parameters can be determined directly from the slopes of the acoustic branches at $k=0$ of the Bloch-Floquet dispersion diagram. To this end, we introduce
\begin{equation}
    c_{\rm p}^{(\phi)}
    \coloneqq
    \left.\frac{\intd\omega_{\rm p}^{(\phi)}}{\intd k}\right|_{k=0},
    \qquad
    c_{\rm s}^{(\phi)}
    \coloneqq
    \left.\frac{\intd\omega_{\rm s}^{(\phi)}}{\intd k}\right|_{k=0},
    \qquad
    \phi\in\left\{0^\circ,45^\circ\right\},
\end{equation}
where $\omega_{\rm p}^{(\phi)}$ and $\omega_{\rm s}^{(\phi)}$ denote
the acoustic pressure and shear branches, respectively. For tetragonal
symmetry, the corresponding wave speeds satisfy
\begin{align}
    \left(c_{\rm p}^{(0)}\right)^2
    &=
    \frac{\kappa_{\rm M}+\mu_{\rm M}}{\rho}\,,
    &
    \left(c_{\rm s}^{(0)}\right)^2
    &=
    \frac{\mu_{\rm M}^{*}}{\rho}\,,
    \notag\\
    \left(c_{\rm p}^{(45)}\right)^2
    &=
    \frac{\kappa_{\rm M}+\mu_{\rm M}^{*}}{\rho}\,,
    &
    \left(c_{\rm s}^{(45)}\right)^2
    &=
    \frac{\mu_{\rm M}}{\rho}\,.
    \label{voss.eq:macro_wave_speeds}
\end{align}
Consequently,
\begin{align}
    \mu_{\rm M}
    &=
    \rho\left(c_{\rm s}^{(45)}\right)^2,
    \hspace{2.8cm}
    \mu_{\rm M}^{*}
    =
    \rho\left(c_{\rm s}^{(0)}\right)^2,
    \notag\\
    \kappa_{\rm M}
    &=
    \rho\left[\left(c_{\rm p}^{(0)}\right)^2-\left(c_{\rm s}^{(45)}\right)^2\right]
    =
    \rho\left[\left(c_{\rm p}^{(45)}\right)^2-\left(c_{\rm s}^{(0)}\right)^2\right].
    \label{voss.eq:macro_parameters}
\end{align}
The last equality provides a consistency relation between the acoustic slopes along the two symmetry directions.

The macroscopic tensor is not an additional independent constitutive tensor of the relaxed micromorphic model. Instead, it is related to the mesoscopic and microscopic elasticity tensors \cite{neff2020identification}.
To see this, we consider the static homogeneous limit of the second Euler-Lagrange equation, for which all inertia and curvature contributions vanish. Hence, the Euler-Lagrange equation simplify to $\widetilde{\sigma}=s$, i.e.\ its symmetric part gives
\begin{equation}
    \mathbb{C}_{\rm e}\.\sym(\nabla u-P)
    =
    \mathbb{C}_{\rm micro}\.\sym P.
\end{equation}
Therefore,
\begin{equation}
    \sym P
    =
    \left(\mathbb{C}_{\rm e}+\mathbb{C}_{\rm micro}\right)^{-1}\mathbb{C}_{\rm e}\.\sym\nabla u
\end{equation}
and the corresponding macroscopic stress can be written as
\begin{equation}
    \widetilde{\sigma}
    =
    \mathbb{C}_{\rm macro}\.\sym\nabla u\,,
    \qquad\qquad
    \mathbb{C}_{\rm macro}
    =
    \mathbb{C}_{\rm micro}\left(\mathbb{C}_{\rm micro}+\mathbb{C}_{\rm e}\right)^{-1}\mathbb{C}_{\rm e}\,.
    \label{voss.eq:Cmacro}
\end{equation}
Equivalently,
\begin{equation}
    \mathbb{C}_{\rm e}
    =
    \mathbb{C}_{\rm micro}
    \left(\mathbb{C}_{\rm micro}-\mathbb{C}_{\rm macro}\right)^{-1}\mathbb{C}_{\rm macro}\,.
    \label{voss.eq:Ce_macro_micro}
\end{equation}
For the present tetragonal parametrization, this relation reduces to
\begin{equation}
    \kappa_{\rm e}
    =
    \frac{\kappa_{\rm m}\kappa_{\rm M}}{\kappa_{\rm m}-\kappa_{\rm M}}\,,
    \qquad\qquad
    \mu_{\rm e}
    =
    \frac{\mu_{\rm m}\mu_{\rm M}}{\mu_{\rm m}-\mu_{\rm M}}\,,
    \qquad\qquad
    \mu_{\rm e}^{*}
    =
    \frac{\mu_{\rm m}^{*}\mu_{\rm M}^{*}}{\mu_{\rm m}^{*}-\mu_{\rm M}^{*}}\,.
    \label{voss.eq:Ce_scalar}
\end{equation}
Thus, after identifying $\{\kappa_{\rm M},\mu_{\rm M},\mu_{\rm M}^{*}\}$ from the acoustic slopes, the three mesoscopic parameters $\{\kappa_{\rm e},\mu_{\rm e},\mu_{\rm e}^{*}\}$ are no longer independent once $\{\kappa_{\rm m},\mu_{\rm m},\mu_{\rm m}^{*}\}$ are prescribed.

\subparagraph{Cut-off frequencies:}
A second set of analytical relations follows from the non-zero cut-off frequencies of the optical branches.
Setting $k=0$ in the generic dispersion relation \eqref{voss.eq:dispRelationAll} gives
\begin{equation}
    -c_2\.\omega^2+c_5\.\omega^4-c_8\.\omega^6=0
    \qquad\iff\qquad
    \omega^2\left(c_2-c_5\.\omega^2+c_8\.\omega^4\right)=0\,.
\end{equation}
The vanishing solution corresponds to the acoustic branch always starting at the origin $\omega=0$, whereas the two non-zero roots correspond to the optical cut-offs.
Since all terms containing spatial derivatives vanish for $k=0$, these cut-offs are independent of the curvature tensors $\mathbb{L}_{\rm s}$, $\mathbb{L}_{\rm a}$, the gradient micro-inertia tensors $\mathbb{T}_{\rm e}$, $\mathbb{T}_{\rm c}$, and the curvature micro-inertia tensors $\mathbb{M}_{\rm s}$, $\mathbb{M}_{\rm a}$.

For tetragonal symmetry, the four characteristic non-zero cut-off frequencies can be written as
\begin{equation}
    \omega_{\rm p}
    =
    \sqrt{\frac{\kappa_{\rm e}+\kappa_{\rm m}}{\rho L_{\rm c}^2\kappa_\gamma}}\,,
    \qquad
    \omega_{\rm s}
    =
    \sqrt{\frac{\mu_{\rm e}+\mu_{\rm m}}{\rho L_{\rm c}^2\gamma_1}}\,,
    \qquad
    \omega_{\rm ss}
    =
    \sqrt{\frac{\mu_{\rm e}^{*}+\mu_{\rm m}^{*}}{\rho L_{\rm c}^2\gamma_1^*}}\,,
    \qquad
    \omega_{\rm r}
    =
    \sqrt{\frac{\mu_{\rm c}}{\rho L_{\rm c}^2\gamma_2}}\,.
    \label{voss.eq:cutoff}
\end{equation}
These cut-off frequencies are independent of the propagation direction $\phi$.
Indeed, because a cut-off is obtained at $k=0$, all terms involving spatial derivatives vanish.
Consequently, the wave vector $\boldsymbol{k}=k\khat(\phi)$ including its directional information $\khat(\phi)$ disappears completely.
Their occurrence on the pressure and shear branches along the two principal symmetry directions is summarized by:
\begin{center}
    \renewcommand{\arraystretch}{\stretch}
    \begin{tabular}{c||c|c}
       cut-offs & $\phi=0^\circ$ & $\phi=45^\circ$\\
        \hline
        pressure & $\omega_{\rm s},\ \omega_{\rm p}$
                 & $\omega_{\rm ss},\ \omega_{\rm p}$\\
        shear    & $\omega_{\rm ss},\ \omega_{\rm r}$
                 & $\omega_{\rm s},\ \omega_{\rm r}$
    \end{tabular}
\end{center}
Hence, once the elastic parameters $\{\kappa_{\rm m},\mu_{\rm m},\mu_{\rm m}^{*},\mu_{\rm c}\}$ are fixed, the four free $\{\kappa_\gamma,\gamma_1,\gamma_1^*,\gamma_2\}$ micro-inertia parameters can be eliminated analytically according to
\begin{equation}
    \kappa_\gamma
    =
    \frac{\kappa_{\rm e}+\kappa_{\rm m}}{\rho L_{\rm c}^2\omega_{\rm p}^2}\,,
    \qquad
    \gamma_1
    =
    \frac{\mu_{\rm e}+\mu_{\rm m}}{\rho L_{\rm c}^2\omega_{\rm s}^2}\,,
    \qquad
    \gamma_1^*
    =
    \frac{\mu_{\rm e}^{*}+\mu_{\rm m}^{*}}{\rho L_{\rm c}^2\omega_{\rm ss}^2}\,,
    \qquad
    \gamma_2
    =
    \frac{\mu_{\rm c}}{\rho L_{\rm c}^2\omega_{\rm r}^2}\,.
    \label{voss.eq:micro_inertia_cutoffs}
\end{equation}
If the Bloch-Floquet results are reported in terms of the ordinary frequency $f$ rather than the angular frequency $\omega$, the corresponding values in \eqref{voss.eq:cutoff} have to be converted according to $\omega=2\pi f$.

Last, the four non-zero cut-off frequencies observed in the Bloch-Floquet spectrum must subsequently be associated with the theoretical quantities $\omega_p,\omega_s,\omega_{ss},\omega_r$.
Typically, we observe a superimposed cut-off of one shear and pressure curve which gives us $\omega_s=\omega_{ss}$.
In addition, the first optic curve is always always a shear mode which we identify as $\omega_r$.

\subparagraph{Asymptotic frequencies:}
Further analytical information can be obtained from the horizontal asymptotes of the dispersion curves.
The occurrence of horizontal asymptotes follows directly from the polynomial structure of the dispersion relation.
Indeed, after division of \eqref{voss.eq:dispRelationAll} by $k^4$, all terms of lower order in $k$ vanish for $k\to\infty$ and for $\omega(k)\to\omega_\infty<\infty$ its limiting frequency necessarily satisfies
\begin{equation}
    c_1-c_4\.\omega_\infty^2+c_7\.\omega_\infty^4-c_{10}\.\omega_\infty^6=0\,.\label{voss.eq:asymptote_polynomial}
\end{equation}
We restate that the coefficients $c_i$
depend on wave type (shear or pressure) and direction of propagation $\phi=0^\circ,45^\circ$.
The limiting polynomial \eqref{voss.eq:asymptote_polynomial} therefore defines a horizontal asymptote of the corresponding continuum branch.

For the plane-strain constitutive structure introduced above, the
curvature tensors are characterized by the single parameters
$\alpha_1$ and $\beta_1$, respectively. In this case, the 
asymptotic frequency polynomial \eqref{voss.eq:asymptote_polynomial} factorizes according to
\begin{equation}
    c_1-c_4\.\omega_\infty^2+c_7\.\omega_\infty^4-c_{10}\.\omega_\infty^6
    = \left(1-\rho L_{\rm c}^2\frac{\beta_1}{\alpha_1}\omega_\infty^2\right)\left(c_1-c_4^{\rm red}\.\omega_\infty^2+c_7^{\rm red}\.\omega_\infty^4\right)
    =0\,,
    \label{voss.eq:asymptote_factorization}
\end{equation}
where $c_4^{\rm red}$ and $c_7^{\rm red}$ denote the corresponding coefficients obtained without the $\Curl\dot P$ contribution \cite{voss2023modeling}.
Hence, the three asymptotic frequencies are given by
\begin{equation}
    \left(\omega_{\infty,1/2}\right)^2
    =\frac{c_4^{\rm red}\mp\sqrt{\left(c_4^{\rm red}\right)^2-4\.c_1c_7^{\rm red}}}{2\.c_7^{\rm red}}
    \qquad\text{and}\qquad
    \omega_{\infty,3}
    =\sqrt{\frac{\alpha_1}{\rho L_{\rm c}^2\.\beta_1}}\,.
    \label{voss.eq:asymptotes}
\end{equation}
The two asymptotic frequencies $\omega_{\infty,1/2}$ depend on the wave type and the direction of propagation and therefore provide eight asymptotic values in total which are, unfortunately, to lengthy to give them here.

In contrast, the third asymptote $\omega_{\infty,3}$ is independent of both the wave type and the propagation direction. This degeneracy is a consequence of the isotropic form retained by $\Curl P$ and $\Curl\dot P$ under the plane-strain assumption.
Indeed, under the plane-strain assumption
\begin{equation}
    P=\matr{P_{11}&P_{12}&0\\P_{21}&P_{22}&0\\0&0&0}
    \qquad\implies\qquad
    \Curl P
    =\matr{0&0&P_{12,1}-P_{11,2}\\
        0&0&P_{22,1}-P_{21,2}\\
        0&0&0}.
\end{equation}
The same structure holds for $\Curl\dot P$.
Thus, only two components of the row-wise Curl remain active in plane strain and, with the constitutive restrictions $\alpha_2=\alpha_1$ and $\beta_2=\beta_1$, the curvature and curvature-inertia contributions are each governed by a single scalar parameter.
This explains why the additional asymptote $\omega_{\infty,3}$ is identical for pressure and shear waves and for both principal propagation directions.

After the preceding identification of the macroscopic elastic parameters and the cut-off frequencies, the eight other asymptotes depend on the four remaining microscopic elastic parameters
$\{\kappa_{\rm m},\mu_{\rm m},\mu_{\rm m}^{*},\mu_{\rm c}\}$
and on three independent combinations of the gradient micro-inertia parameters $\{\overline{\kappa}_1,\overline{\kappa}_2,\overline{\kappa}_3\}$. 
Thus, seven bulk-relevant parameters remain to be identified from eight asymptotic values.
We collect them in
\begin{equation}
    p_\infty\left(\kappa_{\rm m},
        \mu_{\rm m},
        \mu_{\rm m}^{*},
        \mu_{\rm c},
        \overline{\kappa}_1,
        \overline{\kappa}_2,
        \overline{\kappa}_3
    \right).
    \label{voss.eq:asymptote_parameters}
\end{equation}
In practice, the continuum asymptotes are matched to the corresponding high-wavenumber characteristics of the Bloch-Floquet branches and the overdetermined system is solved in a least-squares sense.
We explain more of the numerically fitting procedure in Section, \ref{voss.sec:numericalFitting}.

Once these seven parameters have been identified, the common third asymptote determines the ratio of the two curvature parameters,
\begin{equation}
    \frac{\alpha_1}{\beta_1}
    =
    \rho L_{\rm c}^2\omega_{\infty,3}^2
    \qquad\iff\qquad
    \beta_1=\frac{\alpha_1}{\rho L_{\rm c}^2\omega_{\infty,3}^2}\,.
    \label{voss.eq:alpha_beta_ratio}
\end{equation}
Consequently, we use the common asymptotes to determine only $\beta_1$ while $\alpha_1$ remains arbitrary.
One additional parameter therefore remains available to adjust the shape of the dispersion curves at finite wavenumbers.

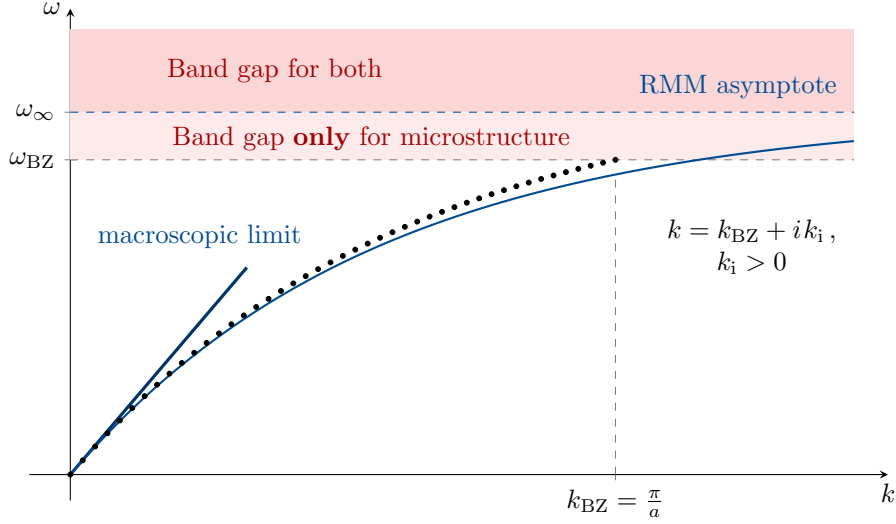
\begin{figure}[h!]
    \centering
    \begin{tikzpicture}
        \begin{axis}[
            axis x line=middle, axis y line=middle,
            x label style={at={(current axis.right of origin)},anchor=north},
            y label style={at={(current axis.above origin)},anchor=east},
            xlabel=$k$, ylabel=$\omega$,
            xmin=-0.6, xmax=12,
            ymin=-0.25, ymax=4.5,
            width=0.8\linewidth,
            height=0.5\linewidth,
            xtick={8},
            xticklabels={$k_{\rm BZ}=\frac{\pi}{a}$},
            ytick={3.04,3.5},
            yticklabels={$\omega_{\rm BZ}$,$\omega_\infty$},
            samples=100,
            clip=false
        ]

        \fill[red!8]  (axis cs:0,3.04) rectangle (axis cs:11.5,3.5);
        \fill[red!16] (axis cs:0,3.5) rectangle (axis cs:11.5,4.3);

        \addplot[gray,dashed] coordinates {(0,3.04) (11.5,3.04)};
        \addplot[blue,dashed] coordinates {(0,3.5) (11.5,3.5)};
        \addplot[gray,dashed] coordinates {(8,0) (8,2.9)};
        \addplot[blue!70!black,very thick] coordinates {(0,0) (2.59,2)};

        \addplot[blue,thick,smooth,domain=0:11.5]{3.5*(1-exp(-0.22*x))};

        \addplot[only marks,mark=*,mark options={solid,black,scale=0.45},domain=0:8,samples=45]
            {3.5*(1-exp(-0.22*x))+0.02*(1/(x+1)+x-1)};

        \node[blue] at (axis cs:1.9,2.3) {macroscopic limit};
        \node[blue] at (axis cs:9.8,3.75) {RMM asymptote};
        \node[red!70!black] at (axis cs:4.42,3.25)
            {Band gap \textbf{only} for microstructure};
        \node[red!70!black] at (axis cs:3.0,3.9)
            {Band gap for both};
        \node[align=center] at (axis cs:10,2.2)
            {$k=k_{\rm BZ}+i\.k_{\rm i}\,,$\\ $k_{\rm i}>0$};
        \end{axis}
    \end{tikzpicture}
    \caption{Schematic comparison of the Bloch-Floquet acoustic branch (black dots) up to $\omega_{\rm BZ}$ and the corresponding RMM branch (blue).
    Both coincide in the same macroscopic limit, while only the RMM continues beyond the Brillouin-zone boundary towards its asymptote $\omega_\infty>\omega_{\rm BZ}$.}
    \label{voss.fig:blochRMM_asymptote}
\end{figure}

It should be emphasized that the limit $k\to\infty$ is a property of the effective continuum model and has no direct counterpart in a Bloch-Floquet computation.
For a periodic microstructure, it is sufficient to consider wavevectors within the bounded Brillouin zone.
Consequently, Bloch-Floquet data are only available up to a finite, direction-dependent wavenumber $k_{\rm BZ}$.
For a square unit cell of size $a$ the first Brillouin-zone boundary is reached at $k_{\rm BZ}=\pi/a$ along $0^\circ$ and at $k_{\rm BZ}=\sqrt{2}\pi/a$ along $45^\circ$.
Hence, the continuum asymptotes cannot be measured directly from the Bloch-Floquet spectrum but must be related to the high-wavenumber behavior of the associated branches, cf.\ Figure \ref{voss.fig:blochRMM_asymptote}.

If the acoustic branch reaches a maximal frequency $\omega_{\rm BZ}$ at the boundary of the first Brillouin zone, increasing the excitation frequency of a sufficiently large periodic specimen beyond $\omega_{\rm BZ}$ does not lead to a propagating wave with $k>k_{\rm BZ}$.
Instead, if no other Bloch-Floquet branch exists in this frequency range, the Bloch wavenumber becomes complex,
$k=k_{\rm r}+i\.k_{\rm i}$, where the real part can be chosen within the first Brillouin zone, $|k_{\rm r}|\leq k_{\rm BZ}$, while $k_{\rm i}>0$ describes the spatial attenuation of the wave.
The corresponding wave field contains the factor
$e^{i\.k_{\rm r}\.x}e^{-k_{\rm i}\.x}$ and is therefore evanescent rather than propagating.
For a band gap opening at the Brillouin-zone boundary, one typically has $k_{\rm r}=k_{\rm BZ}$ and $k_{\rm i}>0$.
The frequency interval above $\omega_{\rm BZ}$ and below the next propagating branch therefore constitutes a band gap for the considered propagation direction and wave type.

\subparagraph{Null-Lagrangian of the gradient micro-inertia:}
We recall the existence of a null-Lagrangian in the gradient micro-inertia term $\nabla\dot u$ regarding the parameters $\mathbb{T}_{\rm e},\mathbb{T}_{\rm c}\sim\{\overline{\kappa}_\gamma,\overline{\gamma}_1,\overline{\gamma}_1^*,\overline{\gamma}_2\}$. As shown in Section \ref{voss.sec:kernel} the contribution of the gradient micro-inertia to the bulk equation, and therefore also the dispersion relations, depends only on the three independent parameters $\overline{\kappa}_1$, $\overline{\kappa}_2$, and $\overline{\kappa}_3$, while we have a free kernel parameter $\overline{\kappa}_4$ only affecting the natural boundary traction.

While the exact value of $\overline{\kappa}_4=\overline{\kappa}_\gamma-\overline{\gamma}_1-\overline{\gamma}_1^*+\overline{\gamma}_2$ must be discussed in the setting of finite-size specimen, in the present work, we remove this otherwise undetermined boundary contribution by selecting the representative for which $\overline{\kappa}_4=0$.
This choice does not modify any bulk equation or dispersion relation, but guarantees that the traction contribution associated solely with the null-Lagrangian vanishes identically.

More explicitly, the family
\begin{equation}
    \matr{\overline{\kappa}_\gamma\\ \overline{\gamma}_1\\ \overline{\gamma}_1^*\\ \overline{\gamma}_2}
    =\matr{\overline{\kappa}_\gamma^{\,0}\\ \overline{\gamma}_1^{\,0}\\ \overline{\gamma}_1^{\star,0}\\ \overline{\gamma}_2^{\,0}}
    +\chi\matr{1\\-1\\-1\\1}
    \label{voss.eq:kernel_direction}
\end{equation}
satisfies
\begin{equation}
    \Delta\overline{\kappa}_1
    =\Delta\overline{\kappa}_2
    =\Delta\overline{\kappa}_3
    =0\,,
    \qquad
    \Delta\overline{\kappa}_4=4\.\chi\,.
\end{equation}
Therefore, for any representative of the same bulk material, the choice $\chi=-\frac14\overline{\kappa}_4^{\,0}$ gives $\overline{\kappa}_4=0$.
Equivalently, inversion of the transformation \eqref{voss.eq:gradient_inertia_combinations} gives
\begin{align}
    \overline{\kappa}_\gamma
    &=
    \frac12\.\overline{\kappa}_1+\frac14\.\overline{\kappa}_3+\frac14\.\overline{\kappa}_4\,,
    &\overline{\gamma}_1
    &=
    \frac12\.\overline{\kappa}_1-\frac14\.\overline{\kappa}_3-\frac14\.\overline{\kappa}_4\,,\notag\\
    \overline{\gamma}_1^*
    &=
    \frac12\.\overline{\kappa}_2+\frac14\.\overline{\kappa}_3-\frac14\.\overline{\kappa}_4\,,
    &\overline{\gamma}_2
    &=
    \frac12\.\overline{\kappa}_2 -\frac14\.\overline{\kappa}_3+\frac14\.\overline{\kappa}_4\,.
    \label{voss.eq:gradient_inertia_reconstruction}
\end{align}
For the representative with vanishing kernel contribution $\overline{\kappa}_4=0$, these relations uniquely reconstruct the four original gradient micro-inertia parameters from $\overline{\kappa}_1$, $\overline{\kappa}_2$, and $\overline{\kappa}_3$.

Importantly, this does not require $\overline{\gamma}_2=0$. Hence, the skew-symmetric gradient micro-inertia $\mathbb{T}_{\rm c}$ is retained; only the null-Lagrangian part which is invisible to the bulk dispersion is removed.

The admissibility of the reconstructed gradient micro-inertia is checked by requiring the corresponding coefficients $\overline{\kappa}_\gamma,\overline{\gamma}_1,\overline{\gamma}_1^*,\overline{\gamma}_2$ to remain positive.
After fixing the kernel by $\overline{\kappa}_4=0$, positive definiteness of the active gradient micro-inertia is equivalent to
\begin{equation}
    2\.\overline{\kappa}_1+\overline{\kappa}_3>0\,,\qquad
    2\.\overline{\kappa}_1-\overline{\kappa}_3>0\,,\qquad
    2\.\overline{\kappa}_2+\overline{\kappa}_3>0\,,\qquad
    2\.\overline{\kappa}_2-\overline{\kappa}_3>0\,,
\end{equation}
which can be simplified to
\begin{equation}
    \overline{\kappa}_1>\frac12\.\abs{\overline{\kappa}_3}\,,
    \qquad
    \overline{\kappa}_2>\frac12\.\abs{\overline{\kappa}_3}\,.
    \label{voss.eq:gradient_inertia_positive}
\end{equation}

%
%
%
\subsection{Numerical fitting of the remaining parameters}\label{voss.sec:numericalFitting}
The analytical asymptote fit introduced above first determines the seven-dimensional parameter vector
\begin{equation}
    p_\infty\coloneqq\left(\kappa_{\rm m},\mu_{\rm m},\mu_{\rm m}^{*},\mu_{\rm c},\overline{\kappa}_1,\overline{\kappa}_2,\overline{\kappa}_3\right).
\end{equation}
In addition, because of the common third asymptote determining only $\beta_1$ through \eqref{voss.eq:alpha_beta_ratio}, the additional curvature parameter $\alpha_1$ remains free.
Therefore, the complete parameter vector used for the subsequent fit of the dispersion curves is $p\coloneqq(p_\infty,\alpha_1)\in\mathcal{P}$
where $\mathcal{P}$ denotes the admissible parameter set.

The asymptotic fit provides an initial estimate for $p_\infty$, whereas the possible subsequent full-curve optimization refines these seven parameters together with $\alpha_1$ using the Bloch-Floquet data over the entire admissible wavenumber range.

The mesoscopic elastic parameters follow from \eqref{voss.eq:Ce_scalar}, the free micro-inertia parameters follow from the cut-off relations \eqref{voss.eq:micro_inertia_cutoffs}, the original gradient micro-inertia parameters are reconstructed from \eqref{voss.eq:gradient_inertia_reconstruction}, and the second curvature parameter is fixed by the asymptote \eqref{voss.eq:alpha_beta_ratio}.
Thus, every choice of $p$ uniquely defines all bulk constitutive tensors entering the dispersion relation.

The remaining parameters $p$ are identified numerically from the Bloch-Floquet dispersion curves of the underlying periodic microstructure.
Let
\begin{equation}
    \omega^{\rm BF}_{r,\phi}(k_i)\,,
    \qquad
    r\in\{1,2,3\}\,,
    \qquad
    \phi\in\{0^\circ,45^\circ\}\,,
\end{equation}
denote the three Bloch-Floquet frequencies of either the pressure or shear family at the sampled wavenumber $k_i$, and let $\omega^{\rm RMM}_{r,\phi}(k_i;p)$ denote the corresponding positive roots of the relaxed micromorphic dispersion relation.
We define the normalized least-squares error
\begin{equation}
    \mathcal{E}(p)
    =
    \sum_{\xi\in\{\mathrm{p}0,\mathrm{s}0,\mathrm{p}45,\mathrm{s}45\}}\sum_{r=1}^{3}\sum_{i=1}^{N_k}
    \left(\frac{\omega^{\rm RMM}_{r,\xi}(k_i;p)-\omega^{\rm BF}_{r,\xi}(k_i)}{\omega^{\rm BF}_{r,\xi}(k_i)}\right)^2,
    \label{voss.eq:dispersion_fit_error}
\end{equation}
The identified parameter vector is then obtained from
\begin{equation}
    p_{\rm opt}=\argmin_{p\in\mathcal{P}}\.\mathcal{E}(p)\,.
    \label{voss.eq:dispersion_fit}
\end{equation}
The admissible set $\mathcal{P}$ is restricted such that all relevant constitutive tensors remain positive definite and thus the analytical relations derived above remain well-defined.

%
%
\subsection{Conceptual limitation of the continuum model}
\label{voss.sec:conceptualLimitations}
Before discussing the quality of such a fit, it is important to distinguish limitations caused by the values of the constitutive parameters from limitations that are intrinsic to the mathematical structure of the enriched continua model.
Along the principal symmetry directions, the pressure and shear problems decouple and both dispersion relations have the form \eqref{voss.eq:dispRelationAll}.
Introducing $q\coloneqq k^2$ and $z\coloneqq\omega^2$, this relation can equivalently be written as
\begin{equation}
    p(q,z)=a_2(z)\.q^2+a_1(z)\.q+a_0(z)=0\,,
    \label{voss.eq:dispersion_quadratic_k}
\end{equation}
where
\begin{align}
    a_0(z)&=-c_2\.z+c_5\.z^2-c_8\.z^3,
    \qquad
    a_1(z)=c_0-c_3\.z+c_6\.z^2-c_9\.z^3,\\
    a_2(z)&=c_1-c_4\.z+c_7\.z^2-c_{10}\.z^3.
\end{align}
This representation immediately reveals several qualitative properties of the model.

As already noted in the plane-wave analysis, for a prescribed wavenumber $\overline k$ the dispersion relation is a cubic polynomial in $z=\omega^2$.
Hence, each decoupled wave family possesses three dispersion branches $\omega_1(\overline k)$, $\omega_2(\overline k)$, and $\omega_3(\overline k)$, corresponding to one acoustic and two optical curves in the present setting.
Considering the pressure and shear blocks together therefore gives six in-plane dispersion branches.
Changing the material parameters can modify their position and shape, but cannot increase their number.

More explicitly, let $q_1(\overline\omega)$ and $q_2(\overline\omega)$ denote the two roots of \eqref{voss.eq:dispersion_quadratic_k} for a prescribed frequency $\overline\omega$.
Every positive real root $q_i>0$ defines a positive real wavenumber $k_i=\sqrt{q_i}$ and therefore a propagating plane wave.
A negative real root gives a purely imaginary wavenumber, while a non-real root gives a complex wavenumber; both cases correspond to evanescent modes and do not define propagating branches in the real $(k,\omega)$-dispersion diagram.
Consequently, there may be zero, one, or two distinct positive real wavenumbers, and therefore zero, one, or two propagating branches of the same wave family (shear or pressure), at a prescribed frequency.

This property is essential for reproducing overlapping frequency ranges of different branches belonging to the same wave family, e.g.\ it allows for the acoustic shear curve to reach higher frequency values than the first optic shear mode. This is often necessary for the shear branches at $45^\circ$, see Figure \ref{voss.fig:voss_simple} and \ref{voss.fig:voss_maze}.

The role of the $k^4$ terms becomes particularly transparent by comparison with a reduced formulation without the curvature contributions $\Curl P$ and $\Curl\dot P$.
In that case, the dispersion relation reduces to
\begin{equation}
    p_{\rm red}(q,z)
    =a_1^{\rm red}(z)\.q+a_0^{\rm red}(z)=0\,,
\end{equation}
and therefore defines at most one positive value of $k$ for every admissible frequency within each pressure or shear problem.
The frequency ranges occupied by the three branches of one wave family must consequently remain disjoint, apart from non-generic degeneracies.
Including the curvature terms raises the polynomial degree from one to two in $k^2$ and removes this restriction.

The polynomial degree in $q=k^2$ is also closely related to the possibility of non-monotone dispersion curves.
Along a regular dispersion branch, implicit differentiation of \eqref{voss.eq:dispersion_quadratic_k} gives
\begin{equation}
    v_{\rm g}\coloneqq\frac{\mathrm d\omega}{\mathrm dk}
    =-\frac{\partial_kp(k^2,\omega^2)}{\partial_\omega p(k^2,\omega^2)}
    =-\frac{k}{\omega}\frac{\partial_qp(q,z)}{\partial_zp(q,z)}\,,
    \label{voss.eq:group_velocity_implicit}
\end{equation}
with
\begin{equation}
    \partial_qp(q,z)=a_1(z)+2\.a_2(z)\.q\,,\qquad
    \partial_zp(q,z)=a_2'(z)\.q^2+a_1'(z)\.q+a_0'(z)\,.
\end{equation}
The sign of the group velocity $v_g$ depends on both derivatives and therefore cannot be inferred from the polynomial degree alone.
However, a regular local extremum of a dispersion branch satisfies $\mathrm d\omega/\mathrm dk=0$ with $\partial_zp\neq0$, and therefore
\begin{equation}
    a_1(z)+2\.a_2(z)\.q=0\,.
\end{equation}
If $a_2(z)\neq0$, this condition can be satisfied at an isolated positive value
\begin{equation}
    q=-\frac{a_1(z)}{2\.a_2(z)}>0\,,
\end{equation}
allowing a dispersion branch to change from increasing to decreasing, or vice versa.
In particular, close to such a local extremum, the same frequency can correspond to two distinct positive wavenumbers.
In contrast, if the dispersion relation is only linear in $q$ it does not have the term $a_2(z)$ and thus a local extremum would require $a_1^{\rm red}(z)=0$.
Together with $p(q,z)=0$, this also implies $a_0^{\rm red}(z)=0$ and therefore represents a non-generic degeneracy.
Thus, the quadratic dependence on $k^2$ permits non-monotone dispersion branches and, in particular, branches with negative group velocity.

The distinction is important for parameter identification. A numerical optimization can move cut-offs, asymptotes and intermediate parts of the dispersion curves and can exploit the full range of shapes permitted by \eqref{voss.eq:dispRelationAll}.
It cannot, however, reproduce a Bloch-Floquet spectrum whose topology is incompatible with the polynomial structure itself. In particular, the present plane-strain RMM cannot represent more than three pressure or three shear branches at a prescribed wavenumber, and just as important no more than two positive wavenumbers of the same wave family at a prescribed frequency, cf.\ Table \ref{voss.table:overview}.
Such discrepancies indicate a limitation of the chosen continuum ansatz rather than a deficiency of the parameter-fitting algorithm.

\begin{table}[h!]
\caption{Relaxed micromorphic model: dispersion face sheet}
\label{voss.table:overview}
\centering
\renewcommand{\arraystretch}{\stretch}
    \begin{tabular}{@{}p{0.25\textwidth}p{0.53\textwidth}p{0.17\textwidth}@{}}
    \toprule
    \textbf{Property} & \multicolumn{2}{l}{\textbf{RMM in plane strain}} \\
    \midrule
    Wave type
        & \multicolumn{2}{l}{pressure $\mid$ shear} \\
    Fixed $k$
        & \multicolumn{2}{l}{$\det B(\omega,k)=0$: cubic in $z=\omega^2$ $\implies$ $3$ branches each} \\
    Fixed $\omega$
        & \multicolumn{2}{p{0.7\textwidth}}{$\det B(\omega,k)=0$: quadratic in $q=k^2$ $\implies$ at most $2$ positive wavenumbers per type} \\
    Dispersion branches
        & \multicolumn{2}{l}{$3+3$ \quad ($1$ acoustic $+\,2$ optical per type)} \\
    Cut-offs
        & \multicolumn{2}{l}{$k=0$ $\implies$ independent of $\phi$} \\
    Asymptotes
        & \multicolumn{2}{l}{$k\to\infty$ $\implies$ generally dependent on $\phi$} \\
    \midrule
    Analytical fit
        & unit cell geometry $\to$ $\rho$, $L_{\rm c}$
        & ($2$ parameters) \\
        & long wave limit $\to$ $\C_{\rm Macro}$
        & ($3$ parameters) \\
        & cut-offs $\to$ $\mathbb{J}_{\rm M}$, $\mathbb{J}_{\rm c}$
        & ($4$ parameters) \\
        & common asymptote $\to$ $\mathbb M$
        & ($1$ parameter) \\
    Numerical fit
        & full curves $+$ $8$ asymptotes $\to$ $p\in\mathcal{P}$
        & ($8$ parameters) \\
    \bottomrule
    \end{tabular}
\end{table}

%
%
%
\section{Results}
\label{voss.sec:results}

\begin{figure}[h!]
    \centering
    \includegraphics[width=0.49\linewidth]{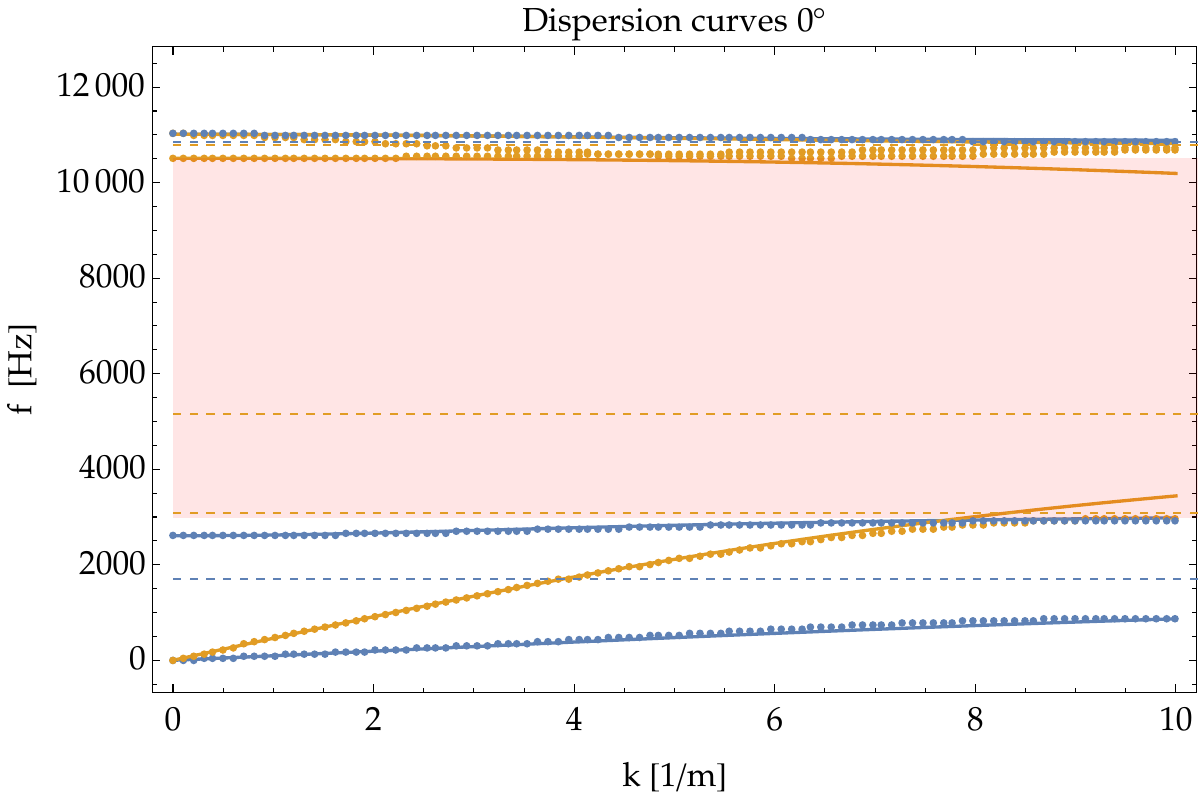}
    \hfill
    \includegraphics[width=0.49\linewidth]{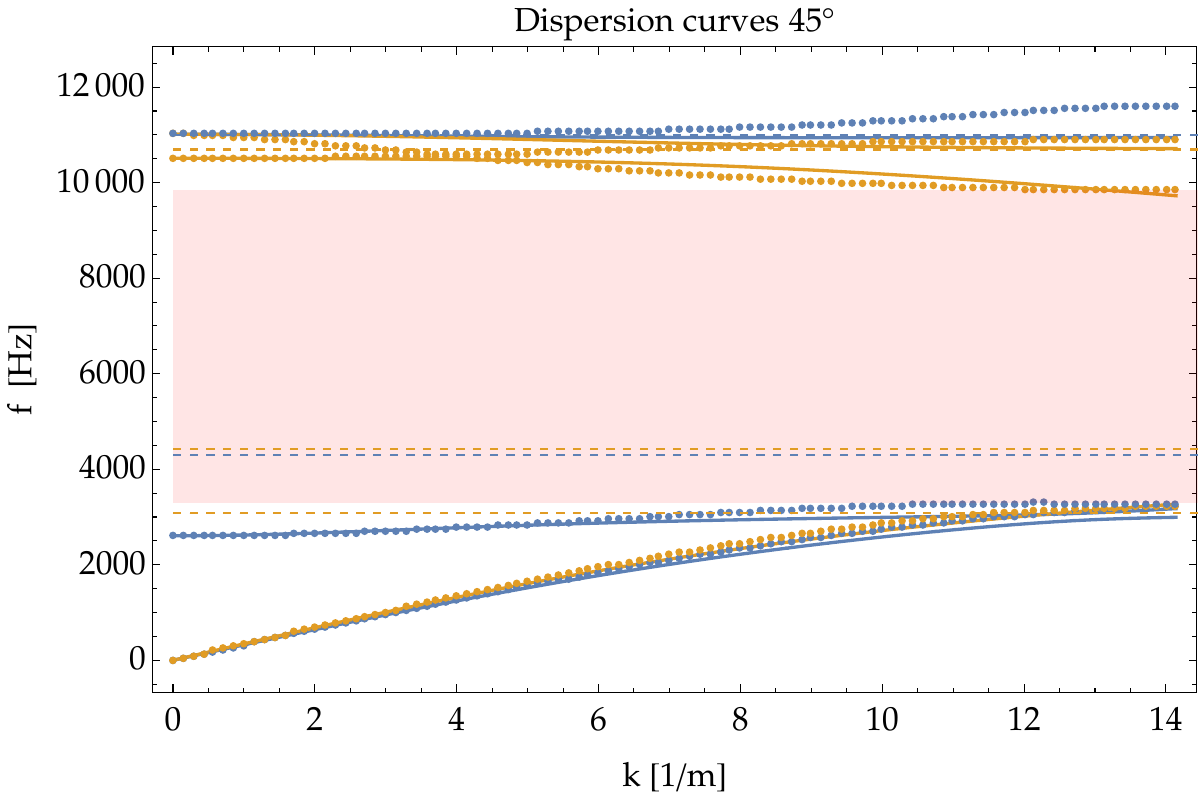}
    \caption{Dispersion frequency curves $f(k)$ for $0^\circ$ (left) and $45^\circ$ (right) with pressure curves colored in yellow and shear in blue. The dots are the points computed with \textit{COMSOL Multiphysics}$^{\circledR}$ while the smooth curves show the analytical expression of the dispersion curves for the relaxed micromorphic model computed with \textit{Mathematica}. The horizontal asymptotes are also shown with dashed lines.}
    \label{voss.fig:fit_simple}
\end{figure}

\begin{figure}[h!]
    \centering
    \includegraphics[width=0.49\linewidth]{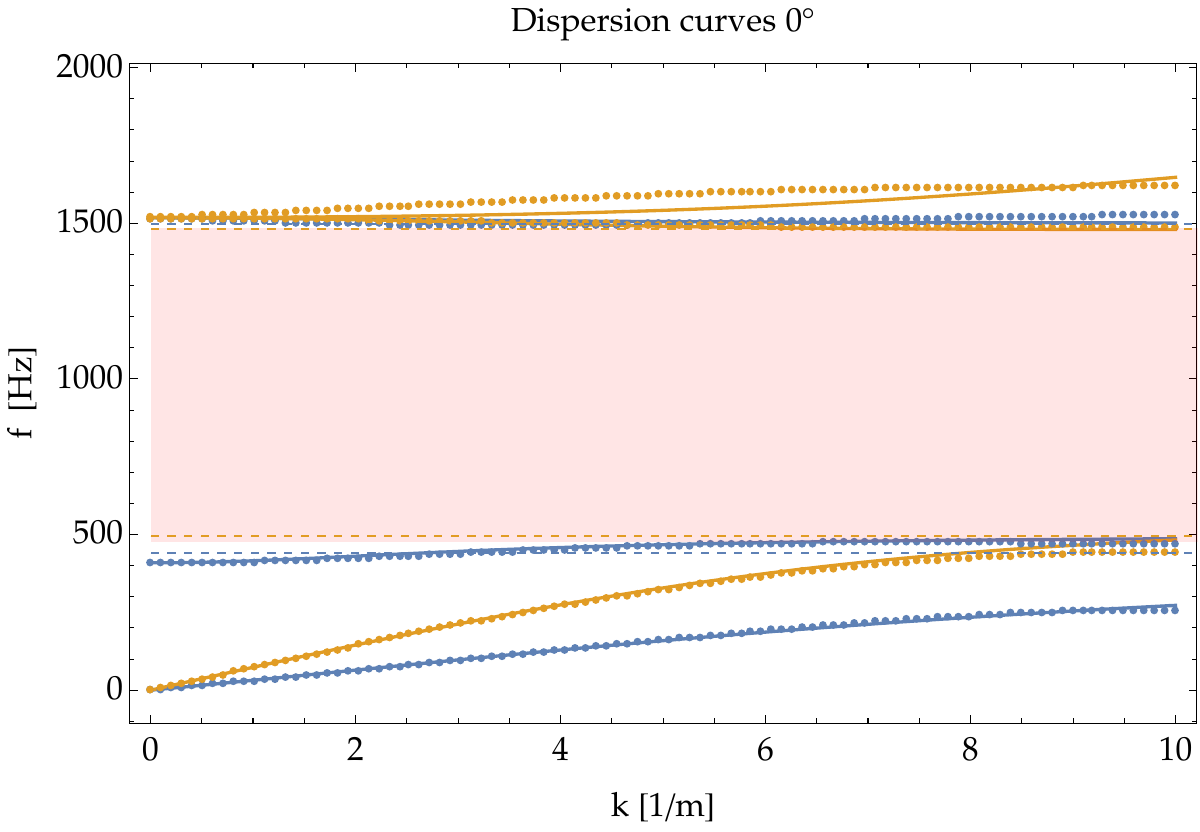}
    \hfill
    \includegraphics[width=0.49\linewidth]{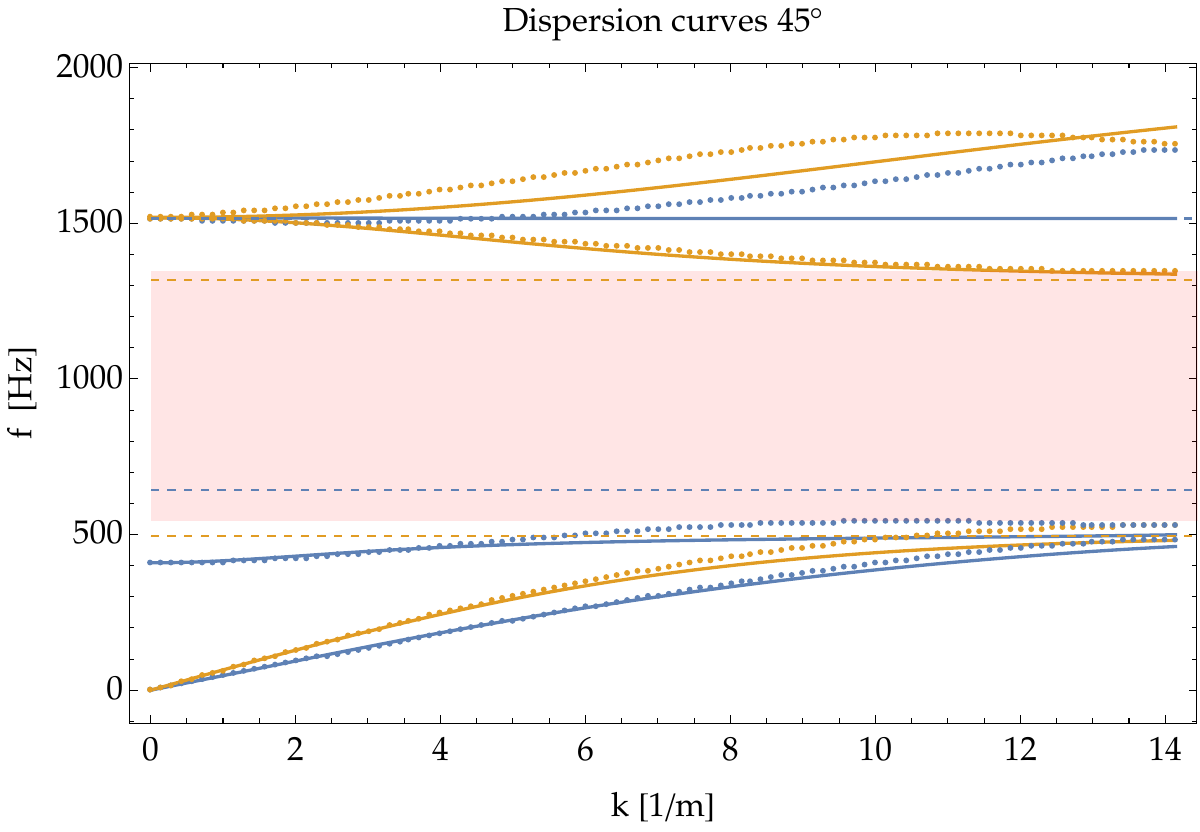}
    \caption{Dispersion frequency curves $f(k)$ for $0^\circ$ (left) and $45^\circ$ (right) with pressure curves colored in yellow and shear in blue. The dots are the points computed with \textit{COMSOL Multiphysics}$^{\circledR}$ while the smooth curves show the analytical expression of the dispersion curves for the relaxed micromorphic model computed with \textit{Mathematica}. The horizontal asymptotes are also shown with dashed lines.}
    \label{voss.fig:fit_maze}
\end{figure}

The identification procedure developed above separates the determination of the relaxed micromorphic parameters into an analytical and a numerical part for the remaining eight-dimensional parameter vector $p\in\mathcal{P}$.
Figures \ref{voss.fig:fit_simple} and \ref{voss.fig:fit_maze} compare the resulting RMM dispersion curves with the Bloch-Floquet curves for propagation along the two symmetry directions $\phi=0^\circ$ and $\phi=45^\circ$ of the periodic unit cells in Figure \ref{voss.fig:voss_simple} and Figure \ref{voss.fig:voss_maze}, respectively.
We choose the common asymptote $\omega_{\infty,3}=\sqrt{\frac{\alpha_1}{\rho L_{\rm c}^2\.\beta_1}}$ at the average of the four associated curves of Bloch-Floquet analysis, i.e.\ one shear and pressure mode each along $0^\circ$ and $45^\circ$.
For both our unit cells this is the case when we choose the acoustic pressure curve together with the first optic shear curve.

The comparison comprises the three pressure and three shear branches for both directions, such that the eight numerically optimized parameters simultaneously control the shape of twelve dispersion curves.
Thus the individual curves are not fitted independently but are coupled through the common set of constitutive parameters (see Table \ref{voss.table:fitting}) and their analytical relations.

The overall fit is good for both unit cells with a normalized mean-square error of $0.0624$ and $0.0394$, respectively.
In particular, the relaxed micromorphic model follows the main decreasing mode above the band gap.
In general, such branches with negative group velocity cannot be by lower-order continuum descriptions whose dispersion relation is only linear in $k^2$, including the reduced relaxed micromorphic formulation without the curvature contributions $\Curl P$ and $\Curl\dot P$.

The overall fit is good for both unit cells with a normalized mean-square error of $0.0624$ and $0.0394$, respectively. In particular, the relaxed micromorphic model follows the main decreasing mode above the band gap, which simpler enriched continuum without the curvature terms $\Curl P$ and $\Curl \dot P$, such as the reduced relaxed micromorphic model, fail to achieve.

\begin{table}[h!]
    \centering
    \caption{Identified material parameters of the relaxed micromorphic model for the simple metamaterial (left) and the maze metamaterial (right) including the macro-limit $\mathbb{C}_{\rm macro}$.}
    \label{voss.table:fitting}
    \renewcommand{\arraystretch}{\stretch}
    \begin{tabular}{c c c c}
    \toprule
    Tensor & Parameter & Value & Unit \\
    \hline
    \multirow{2}{*}{--}
        & $L_c$ & $50$ & $\mathrm{mm}$ \\
        & $\rho$ & $540.7$ & $\mathrm{kg/m^3}$ \\
    \hline
    \multirow{3}{*}{$\mathbb{C}_{\rm micro}$}
        & $\kappa_{\rm m}$ & $1.208\times 10^{9}$ & $\mathrm{Pa}$ \\
        & $\mu_{\rm m}$ & $5.179\times 10^{8}$ & $\mathrm{Pa}$ \\
        & $\mu_{\rm m}^{*}$ & $5.179\times 10^{8}$ & $\mathrm{Pa}$ \\
    $\mathbb{C}_{\rm c}$
        & $\mu_{\rm c}$ & $9.738\times 10^{5}$ & $\mathrm{Pa}$ \\
    \hline
    \multirow{3}{*}{$\mathbb{C}_{\rm e}$}
        & $\kappa_{\rm e}$ & $6.259\times 10^{7}$ & $\mathrm{Pa}$ \\
        & $\mu_{\rm e}$ & $6.389\times 10^{7}$ & $\mathrm{Pa}$ \\
        & $\mu_{\rm e}^{*}$ & $4.909\times 10^{6}$ & $\mathrm{Pa}$ \\
    \hline
    \multirow{3}{*}{$\mathbb{J}_{\rm m}$}
        & $\kappa_{\gamma}$ & $8.511$ & -- \\
        & $\gamma_1$ & $3.549$ & -- \\
        & $\gamma_1^*$ & $3.189$ & -- \\
    $\mathbb{J}_{\rm c}$
        & $\gamma_2$ & $0.1060$ & -- \\
    \hline
    \multirow{3}{*}{$\mathbb{T}_{\rm e}$}
        & $\overline{\kappa}_{\gamma}$ & $1.641$ & -- \\
        & $\overline{\gamma}_1$ & $1.557$ & -- \\
        & $\overline{\gamma}_1^*$ & $0.7904$ & -- \\
    $\mathbb{T}_{\rm c}$
        & $\overline{\gamma}_2$ & $0.7060$ & -- \\
    \hline
    $\mathbb{L}_{\rm s},\,\mathbb{L}_{\rm a}$
        & $\alpha_1$ & $3.213\times 10^{7}$ & $\mathrm{Pa}$ \\
    $\mathbb{M}_{\rm s},\,\mathbb{M}_{\rm a}$
        & $\beta_1$ & $2.487$ & -- \\
    \midrule
    \multirow{3}{*}{$\mathbb{C}_{\rm macro}$}
        & $\kappa_{\rm M}$ & $5.951\times 10^{7}$ & $\mathrm{Pa}$ \\
        & $\mu_{\rm M}$ & $5.687\times 10^{7}$ & $\mathrm{Pa}$ \\
        & $\mu_{\rm M}^{*}$ & $4.863\times 10^{6}$ & $\mathrm{Pa}$ \\
    \bottomrule
    \end{tabular}
    \qquad
    \begin{tabular}{c c c c}
    \toprule
    Tensor & Parameter & Value & Unit \\
    \hline
    \multirow{2}{*}{--}
        & $L_c$ & $50$ & $\mathrm{mm}$ \\
        & $\rho$ & $540.7$ & $\mathrm{kg/m^3}$ \\
    \hline
    \multirow{3}{*}{$\mathbb{C}_{\rm micro}$}
        & $\kappa_{\rm m}$ & $2.092\times 10^{6}$ & $\mathrm{Pa}$ \\
        & $\mu_{\rm m}$ & $5.178\times 10^{8}$ & $\mathrm{Pa}$ \\
        & $\mu_{\rm m}^{*}$ & $7.216\times 10^{7}$ & $\mathrm{Pa}$ \\
    $\mathbb{C}_{\rm c}$
        & $\mu_{\rm c}$ & $1.463\times 10^{5}$ & $\mathrm{Pa}$ \\
    \hline
    \multirow{3}{*}{$\mathbb{C}_{\rm e}$}
        & $\kappa_{\rm e}$ & $1.179\times 10^{7}$ & $\mathrm{Pa}$ \\
        & $\mu_{\rm e}$ & $1.178\times 10^{6}$ & $\mathrm{Pa}$ \\
        & $\mu_{\rm e}^{*}$ & $5.415\times 10^{5}$ & $\mathrm{Pa}$ \\
    \hline
    \multirow{3}{*}{$\mathbb{J}_{\rm m}$}
        & $\kappa_{\gamma}$ & $4.448$ & -- \\
        & $\gamma_1$ & $166.7$ & -- \\
        & $\gamma_1^*$ & $23.35$ & -- \\
    $\mathbb{J}_{\rm c}$
        & $\gamma_2$ & $0.6462$ & -- \\
    \hline
    \multirow{3}{*}{$\mathbb{T}_{\rm e}$}
        & $\overline{\kappa}_{\gamma}$ & $0.9686$ & -- \\
        & $\overline{\gamma}_1$ & $1.036$ & -- \\
        & $\overline{\gamma}_1^*$ & $1.265$ & -- \\
    $\mathbb{T}_{\rm c}$
        & $\overline{\gamma}_2$ & $1.333$ & -- \\
    \hline
    $\mathbb{L}_{\rm s},\,\mathbb{L}_{\rm a}$
        & $\alpha_1$ & $1.229\times 10^{7}$ & $\mathrm{Pa}$ \\
    $\mathbb{M}_{\rm s},\,\mathbb{M}_{\rm a}$
        & $\beta_1$ & $37.27$ & -- \\
    \midrule
    \multirow{3}{*}{$\mathbb{C}_{\rm macro}$}
        & $\kappa_{\rm M}$ & $1.777\times 10^{6}$ & $\mathrm{Pa}$ \\
        & $\mu_{\rm M}$ & $1.175\times 10^{6}$ & $\mathrm{Pa}$ \\
        & $\mu_{\rm M}^{*}$ & $5.374\times 10^{5}$ & $\mathrm{Pa}$ \\
    \bottomrule
    \end{tabular}
\end{table}
We naturally see the largest difference between the two descriptions in their high-wavenumber behavior.
The RMM branches approach horizontal asymptotes $\omega_\infty$ for $k\to\infty$, whereas the Bloch-Floquet branches are bounded by $k_{\rm BZ}$ generating a gap $[\omega(k_{\rm BZ}),\omega_\infty]$.
Consequently, Bloch-Floquet branches that already exhibit a small slope close to the Brillouin-zone boundary $k_{\rm BZ}$ are naturally more compatible with the asymptotic structure of an enriched continua model.
Conversely, a pronounced slope or curvature at $k_{\rm BZ}$ introduces a structural mismatch that cannot be removed by adjusting the constitutive parameters.
A unit-cell design that matches better the necessary asymptotic behavior of the RMM is discussed in \cite{demore2022unfolding}.

%
%
%
\subsection{Further analytical characterization}
\label{voss.sec:furtherAnalytical}
The analytical identification developed above is not exhaustive.
Additional local characteristics of the dispersion curves can, in principle, be related to the constitutive parameters by differentiating the implicit dispersion relation $p(k^2,\omega^2)$ at $k=0$.
For instance, an optical cut-off $k=0$ with $\omega(0)=\omega_{\rm c}>0$, symmetry of the dispersion relation with respect to $k$ implies $\omega'(0)=0$ and therefore contains no additional constitutive information.
However, if we instead consider the curvature $\omega''(0)$ which can be obtained by a second implicit differentiation of the dispersion relation, we will indeed obtain additional constitutive analytical expressions.

The question is therefore not whether further analytical relations can be derived, but whether they provide sufficiently independent information to justify the increasing algebraic complexity.
In addition, such high-order conditions at $k=0$ do not ensure that the resulting dispersion curves have a good fit for the whole Brillouin-zone and correct asymptotes.

A complementary source of information is also obtained from the static response of the metamaterial.
The relaxed micromorphic model, as well as other enriched continua, is not restricted to dynamical problems but can equally be employed in the static regime.

In this case, the kinetic energy $J(\dot{u},\nabla\dot{u},\dot{P},\Curl\dot{P})$ vanishes while the strain energy $W(\nabla u,P,\Curl P)$, and therefore the associated elastic material parameters, remains unchanged.
This provides an additional physical interpretation of the elastic tensors entering the dynamic formulation.
In particular, the role of $\mathbb{C}_{\rm micro}$ as the elastic stiffness associated with the response at small length scales, in contrast to the macroscopic stiffness $\mathbb{C}_{\rm macro}$ obtained in the long-wavelength limit, is discussed in detail in \cite{neff2020identification}.
It is therefore natural to complement the Bloch-Floquet identification considered here by independent static tests.
Combining static and dynamic data may provide additional constraints on the elastic parameters and reduce ambiguities that cannot be resolved from dispersion curves alone; such a combined static-dynamic identification strategy has recently been pursued in \cite{sarhil2026static}.

%
%
%
%
\subsection{Finite-size specimens}
\label{voss.sec:finiteSize}

The parameter identification developed in the previous sections is based on Bloch-Floquet dispersion curves and therefore characterizes the bulk behavior of the periodic metamaterial.
For an infinite periodic medium, the particular choice of the unit cell is ambiguous: equivalent but different unit-cell cuts generate the same periodic microstructure and consequently the same Bloch-Floquet spectrum.
On the other hand, the corresponding enriched continuum must therefore be characterized by the same bulk constitutive parameters, independently of the chosen unit-cell representation.

This equivalence is lost the moment a finite number of unit cells is considered as it is done for all applications.
Different cuts of the same periodic architecture generally produce different microstructures at the external boundary of the specimen.
Consequently, finite specimens that are equivalent from the perspective of the infinite periodic medium may exhibit different responses due to their different boundary microstructures.
A homogeneous enriched continuum identified exclusively from bulk dispersion data cannot distinguish between these specimens, since the microscopic position of the external boundary relative to the periodic architecture is no longer contained in the homogenized bulk description.

This distinction is already visible within the present formulation.
As shown in Section~\ref{voss.sec:kernel}, the parameter $\overline{\kappa}_4$ belongs to a null-Lagrangian contribution of the gradient micro-inertia that does not enter the bulk equations but contributes to the natural boundary traction through \eqref{voss.eq:kernel_traction}.
Bulk parameter identification alone is therefore insufficient to determine all boundary-sensitive quantities of the effective model.

Therefore, a more fundamental problem arises from the general relation between the periodic unit cell and the boundary of a finite specimen:
\begin{itemize}
    \item Consider two finite metamaterial specimens constructed from the same infinite periodic medium but cut at different positions within the unit cell.
    \item Their bulk microstructure, and consequently their Bloch-Floquet dispersion spectrum, are identical.
    \item The corresponding homogeneous enriched continuum identified exclusively from this bulk spectrum must therefore assign the same bulk constitutive parameters to both specimens.
    \item Nevertheless, the two microscopic cuts can produce different material distributions at the boundary and thus lead to local deformation mechanisms that may lead to vastly different reflection, transmission, or resonance behavior.
\end{itemize}
This information cannot be encoded by the bulk constitutive tensors alone.
The difficulty is not specific to the relaxed micromorphic model but is inherent to the enriched continua framework which is a purely bulk effective description: homogenization removes the microscopic position of the external boundary relative to the periodic architecture.
A predictive treatment of finite metastructures therefore additionally requires, an effective description of the boundary or interface response.

%
%
%
%
\subsection{Towards effective boundary modeling}
\label{voss.sec:effectiveBoundaries}

The discussion above suggests that further improvements of enriched-continuum descriptions of finite metamaterials should not necessarily be sought by adding further bulk parameters.
Once the relevant Bloch-Floquet dispersion branches are reproduced with sufficient accuracy, additional bulk constitutive freedom cannot recover information that has been lost through the homogenization of the microscopic boundary.

The null-Lagrangian freedom associated with $\kappa_4$ should be distinguished from the additional boundary modeling considered below.
Changing $\kappa_4$ only selects a different representative of the same bulk constitutive response while modifying the natural boundary traction.
By contrast, effective interface forces or a surface Lagrangian introduce additional constitutive information that is not contained in the bulk Bloch-Floquet identification and is intended to account for the specific microscopic boundary cut.

A natural extension is therefore to supplement the bulk enriched continuum by effective interface forces that account for the interaction between the specific microscopic boundary cut and another homogenized medium as well as free boundary effects.
Such interface corrections have already been investigated for finite-size metamaterials within reduced relaxed micromorphic formulations \cite{ramirez2024effective,demetriou2025effective}, i.e.\ without $\Curl P$ and $\Curl\dot P$ contributions.

An alternative, and potentially more systematic, formulation is the introduction of a surface Lagrangian in addition to the existing bulk Lagrangian.
The corresponding (kinetic) surface energy could contain effective parameters associated specifically with the microscopic boundary layer.
Variation of the combined bulk-surface action would then generate the same bulk Euler-Lagrange equations together with modified natural boundary conditions.
In this setting, specimens with identical bulk microstructure but different unit-cell cuts could share the same bulk constitutive parameters while being distinguished through their effective surface parameters and thus distinguish between the different finite-size specimen.

Such a separation between bulk and surface identification would also provide a natural hierarchy for the enriched-continuum modeling framework.
The Bloch-Floquet dispersion data determines the bulk constitutive response, i.e.\ the bulk parameter of the Euler-Lagrange equations, whereas finite-size reflection, transmission, or resonance data can subsequently be used to identify the additional boundary quantities.

A systematic characterization and identification of these effective boundary quantities remains an important direction for further investigation.
In particular, the relation between a given microscopic unit-cell cut and the corresponding effective surface parameters should be established without relying on a separate numerical optimization for every finite specimen.
Such a development would provide a further step from enriched continua that can (only) accurately reproduce the bulk dispersion of metamaterials towards effective models that remain predictive for finite-size structures including their interfaces.

%
%
%
\section*{Acknowledgement}
Angela Madeo, Svenja Hermann, and Jendrik Voss acknowledge support from the European Commission through the funding of the ERC Consolidator Grant META-LEGO, N$^\circ$ 101001759.
Patrizio Neff acknowledges support in the framework of the DFG-Priority Programme 2256 ``Variational Methods for Predicting Complex Phenomena in Engineering Structures and Materials'', Neff 902/10-1, Project-No.\ 440935806.

\footnotesize
\section{References}
\printbibliography[heading=none]
\small

\begin{appendix}
\section{Material tensors - tetragonal symmetry}\label{voss.app:tensor}

We list all material tensors of the relaxed micromorphic model introduced in Section \ref{voss.sec:RMM} written in Voigt-notation $(\sigma_{11},\sigma_{22},\sigma_{33},\sigma_{23},\sigma_{13},\sigma_{12})$ for the symmetric tensors
\begin{align}
    \mathbb{C}_{\rm e}=
    &\matr{%
    \kappa_{\rm e} + \mu_{\rm e}	& \kappa_{\rm e} - \mu_{\rm e}				& \star & \dots	& 0\\ 
    \kappa_{\rm e} - \mu_{\rm e}	& \kappa_{\rm e} + \mu_{\rm e} & \star & \dots & 0\\
    \star & \star & \star & \dots & 0\\
    \vdots & \vdots	& \vdots & \ddots &\\ 
    0 & 0 & 0 & & \mu_{\rm e}^{*}
    },
    &\mathbb{C}_{\rm micro}= 
    &\matr{%
    \kappa_{\rm m} + \mu_{\rm m}	& \kappa_{\rm m} - \mu_{\rm m}				& \star & \dots	& 0\\ 
    \kappa_{\rm m} - \mu_{\rm m}	& \kappa_{\rm m} + \mu_{\rm m} & \star & \dots & 0\\
    \star & \star & \star & \dots & 0\\
    \vdots & \vdots	& \vdots & \ddots &\\ 
    0 & 0 & 0 & & \mu_{\rm m}^{*}
    },\notag\\
	\mathbb{J}_{\rm m}=\rho\. L_{\rm c}^2
	&\matr{%
	\kappa_\gamma + \gamma_{1} & \kappa_\gamma - \gamma_{1} & \star & \dots & 0\\ 
	\kappa_\gamma - \gamma_{1} & \kappa_\gamma + \gamma_{1} & \star & \dots & 0\\ 
	\star & \star & \star & \dots & 0\\
	\vdots & \vdots & \vdots & \ddots &\\ 
	0 & 0 & 0 & & \gamma^{*}_{1}\\ 
	},
	&\mathbb{T}_{\rm e}=\rho\. L_{\rm c}^2
	&\matr{%
	\kappabar_{\gamma} + \gammabar_{1} & \kappabar_{\gamma} - \gammabar_{1} & \star & \dots	& 0\\ 
	\kappabar_{\gamma} - \gammabar_{1} &   \kappabar_{\gamma} + \gammabar_{1} & \star & \dots & 0\\ 
	\star & \star & \star & \dots & 0\\
	\vdots & \vdots & \vdots & \ddots &\\ 
	0 & 0 & 0 & & \gammabar^{*}_{1}
	},\label{eq:micro_ine_1}\\
    \mathbb{L}_{\rm s}=L_{\rm c}^2
    &\matr{%
    \star                   & \star                   & \star                   & \multicolumn{2}{c}{\dots} & 0     \\
    \star                   & \star                   & \star                   & \multicolumn{2}{c}{\dots} & 0     \\
    \star                   & \star                   & \star                   & \multicolumn{2}{c}{\dots} & 0     \\
    \multirow{2}{*}{\vdots} & \multirow{2}{*}{\vdots} & \multirow{2}{*}{\vdots} & \alpha_1    & 0           & 0     \\
                            &                         &                         & 0           & \alpha_1    & 0     \\
    0                       & 0                       & 0                       & 0           & 0           & \star
    },
    &\mathbb{M}_{\rm s}=\rho \. L_{\rm c}^4
    &\matr{%
    \star                   & \star                   & \star                   & \multicolumn{2}{c}{\dots} & 0     \\
    \star                   & \star                   & \star                   & \multicolumn{2}{c}{\dots} & 0     \\
    \star                   & \star                   & \star                   & \multicolumn{2}{c}{\dots} & 0     \\
    \multirow{2}{*}{\vdots} & \multirow{2}{*}{\vdots} & \multirow{2}{*}{\vdots} & \beta_1    & 0           & 0     \\
                            &                         &                         & 0           & \beta_1    & 0     \\
    0                       & 0                       & 0                       & 0           & 0           & \star
    },\notag
\end{align}
and $(\sigma_{23},\sigma_{13},\sigma_{12})$ for the anti-symmetric tensors
\begin{align}
	\mathbb{C}_{\rm c}&= 
    \matr{%
    \star & 0 & 0\\ 
	0 & \star & 0\\ 
	0 & 0 & 4\.\mu_{\rm c}
    },&
	\mathbb{J}_{\rm c}&=\rho\. L_{\rm c}^2
	\matr{%
	\star & 0 & 0\\ 
	0 & \star & 0\\ 
	0 & 0 & 4\.\gamma_{2}
	},&
	\mathbb{T}_{\rm c}&=\rho\. L_{\rm c}^2
	\matr{%
    \star & 0 & 0\\ 
	0 & \star & 0\\ 
	0 & 0 & 4\.\gammabar_{2}
	},\notag\\
    \mathbb{L}_{\rm a}&=
    L_{\rm c}^2
    \matr{%
    4\.\alpha_2 & 0 & 0\\ 
	0 & 4\.\alpha_2 & 0\\ 
	0 & 0 & \star
    },&
    \mathbb{M}_{\rm a}&=\rho \. L_{\rm c}^4
    \matr{%
    4\.\beta_2 & 0 & 0\\ 
	0 & 4\.\beta_2 & 0\\ 
	0 & 0 & \star
    },
\end{align}
where only the parameters involved under the plane-strain hypothesis are explicitly presented. Thereby, the symbol $\star$ indicates that the specific entry do not intervene under the plane-strain hypothesis.
It is emphasized that the matrix representation of $\mathbb{L}_{\rm s}$, $\mathbb{L}_{\rm a}$, $\mathbb{M}_{\rm s}$, and $\mathbb{M}_{\rm a}$ differs since ``out of plane curvatures'' related to $\Curl P$ and $\Curl\dot P$.
In addition, we set $\alpha_2=\alpha_1$ and $\beta_2=\beta_1$ to ensure purely in-plane components in the equilibrium equation without reducing the number of independent ones, cf.\ \cite[Appendix A]{voss2023modeling} for a more detailed explanation.

We will also use $\mathbb{C}_{\rm macro} = \mathbb{C}_{\rm micro}\left(\mathbb{C}_{\rm micro}+\mathbb{C}_{\rm e}\right)^{-1}\mathbb{C}_{\rm e}$ as the effective stiffness tensor in the long-wavelenght limit satisfying the same tetragonal symmetry as shown here for $\mathbb{C}_{\rm micro}$ and $\mathbb{C}_{\rm e}$ with the associated material parameters $\kappa_{\rm M}$, $\mu_{\rm M}$, and $\mu_{\rm M}^{*}$.
Finally, the gradient micro-inertia contains the one-dimensional null-Lagrangian kernel discussed in Section~\ref{voss.sec:kernel}.
Instead of suppressing the skew-symmetric contribution by setting $\overline{\gamma}_2=0$, we retain the full tensors $\mathbb{T}_{\rm e}$ and $\mathbb{T}_{\rm c}$ and select the representative such that the boundary traction associated solely with the null-Lagrangian vanishes.
\end{appendix}
\end{document}